\documentclass[apj]{emulateapj}

\usepackage{graphicx}
\usepackage{dcolumn}
\usepackage{bm}
\usepackage{amsmath}
\usepackage{txfonts}

\usepackage{color}

\usepackage{natbib}

\newcommand{\qm}[1]{``#1''}
\newcommand{\fromto}{\,${--}$\,}

\newcommand*{\T}{\textstyle}

\def\sss{\scriptscriptstyle}
\def\U{{\sss \!U}}
\def\L{{\sss \!L}}
\def\K{{\sss \!K}}

\def\T{{\sss \!T}}

\def\P{{\sss \!P}}

\def\I{{\sss \!I}}

\def\nur{\nu_\mathrm{r}}

\def\nuL{\nu_\L}
\def\nuU{\nu_\U}
\def\nuK{\nu_\K}

\def\nul{\nuL}
\def\nuu{\nuU}

\def\nuP{\nu_\P}

\def\MS{M_{0}}
\def\B{\mathcal{F}}

\definecolor{gray}{rgb}{.6,.6,.6}
\definecolor{green}{rgb}{0,.6,0}
\definecolor{red}{rgb}{.9,0,0}
\newcommand*{\change}[1]{{{#1}}}

\begin{document}
%
   

\title{\vspace{-8mm}On mass-constraints implied by the relativistic precession model of twin-peak quasi-periodic oscillations in Circinus X-1}   

\journalinfo{ORIGINAL PAPER IN: The Astrophysical Journal, 714:748–757, 2010 May 1} \submitted{}

\shorttitle{ON MASS OF CIRCINUS X-1}
\shortauthors{T\"OR\"OK ET AL.}

 \author{Gabriel T\"or\"ok, Pavel Bakala, Eva \v Sr\' amkov\'a, Zden\v ek Stuchl\'ik, Martin Urbanec}
 
  \affil{Institute of Physics, Faculty of Philosophy and Science, Silesian University in Opava,
               Bezru\v covo n\' am. 13, CZ-746\,01 Opava, Czech Republic, \\www.physics.cz;\\terek@volny.cz,pavel.bakala@fpf.slu.cz,sram\_eva@centrum.cz,zdenek.stuchlik@fpf.slu.cz,martin.urbanec@fpf.slu.cz
              }


\begin{abstract}
\noindent
Boutloukos et al. (2006) discovered twin-peak quasi-periodic oscillations (QPOs) in 11 observations of the peculiar Z-source Circinus X-1. Among several other conjunctions the authors briefly discussed the related estimate of the compact object mass following from the geodesic relativistic precession model for kHz QPOs. Neglecting the neutron star rotation they reported the inferred mass $M_0=2.2\pm0.3M_\odot$. We present a more detailed analysis of the estimate which involves the frame-dragging effects associated with rotating spacetimes. For a free mass we find acceptable fits of the model to data for (any) small \change{dimensionless compact object angular momentum $j=cJ/GM^2$}. Moreover, quality of the fit tends to increase very gently with rising $j$. Good fits are reached when $M\sim M_0[1+0.55(j+j^2)]$. It is therefore impossible to estimate the mass without the independent knowledge of the angular momentum and vice versa. \change{Considering $j$ up to 0.3 the range of the feasible values of mass extends up to $3M_\sun$.  We suggest that similar increase of estimated mass due to rotational effects can be relevant for several other sources.} 
\end{abstract}

\keywords{stars: neutron --- X-rays: binaries}


\section{Introduction}

Quasi-periodic oscillations (QPOs) appear in variabilities of several low-mass X-ray binaries (LMXBs) including those which contain a neutron star (NS). A certain kind of these oscillations, the so-called {kHz} (or high-frequency) QPOs, come often in pairs with frequencies $\nuL$ and $\nuU$ typically in the range $\sim\!50\fromto1300\,\mathrm{Hz}$. This is of the same order as the range of frequencies characteristic for orbital motion close to a compact object. \change{Accordingly, most kHz QPO models involve orbital motion in the inner regions of an accretion disk \citep[see][for a recent review]{Kli:2006:CompStelX-Ray:,Lam-Bou:2007:ASSL:ShrtPerBS}.}

\change{There is a large variety of QPO models related to NS sources (in some but not all cases they are applied to black hole (BH) sources too). Concrete models involve miscellaneous mechanisms of producing the observed rapid variability. One of the first possibilities proposed represents the \qm{\emph{beat-frequency}} model assuming interactions between the accretion disk and spinning stellar surface \citep{alp-sha:1985,lam-etal:1985}. Many other models primarily assume {accretion disk oscillations}. For instance, \emph{non-linear resonance} scenarios suggested by Abramowicz, Klu{\'z}niak and Collaborators \citep{abr-klu:2001,klu-abr:2001,abr-etal:2003b,abr-etal:2003c,hor:2008,hor-etal:2009} are often debated. A set of the later models join the beat frequency idea, \emph{magnetic field} influence and presence of the \emph{sonic-point} \citep[][]{mil-etal:1998a,psa-etal:1999,lam-col:2001,lam-col:2003}. Some of numerous versions of non-linear oscillation models and the late beat frequency models rather fade into the same concept that commonly assumes the NS spin to be important for excitation of the resonant effects \citep[][]{lam-col:2003,klu-etal:2004,pet:2005a,pet:2005b,pet:2005c,mil:2006,klu:2008,stu-etal:2008,muk:2009}. Resonance, influence of the spin and magnetic field play a role also in the ideas discussed by \cite{tit-ken:2002} and \cite{tit:2002}. Another resonances are  accommodated in models assuming \emph{deformed disks} \citep[][]{kat:2007,kat:2008,kat:2009a,kat:2009b,meh-tag:2009}.  Further effects induced in the accreted plasma by the NS magnetic field \citep[\emph{Alph\'en wave model}, ][]{zha:2005,zha-etal:2007a,zha-etal:2007b}, oscillations that arise due to \emph{comptonization of the disk--corona} \citep{lee-mil:1998} or oscillations excited in toroidal disk  \citep{rez-etal:2003,rez:2004,sra:2005,sch-rez:2006,bla-etal:2007,sra-etal:2007,str-sra:2009} are considered as well. At last but not least, already the kinematics of the orbital motion itself provides space for consideration of \qm{\emph{hot-spot-like}} models identifying the observed variability with orbital frequencies. For instance recent works of \cite{cad-etal:2008} and \cite{kos-etal:2009} deal with \emph{tidal disruption} of large accreted inhomogenities. Among the same class of (kinematic) models belongs also the often quoted \qm{\emph{relativistic precession}} (RP) kHz QPO model that is of our attention here.}

\change{The RP model has been proposed in a series of papers by \cite{ste-vie:1998,ste-vie:1999,ste-vie:2002}.} It explains the kHz QPOs as a direct manifestation of modes of relativistic epicyclic motion of blobs arising at various radii $r$ in the inner parts of the accretion disk. The model identifies the lower and upper kHz QPOs with the periastron precession $\nuP$ and Keplerian $\nuK$ frequency,
\begin{equation}
\label{equation:stella}
\nul(r)=\nuP(r)=\nuK(r)-\nur(r),\quad\nuu(r)=\nuK(r),
\end{equation}
where $\nu_{r}$ is the radial epicyclic frequency of the Keplerian motion. \change{Note that, on a formal side, for Schwarzschild spacetime where $\nuK$ equals a vertical epicyclic frequency this identification merges with a model assuming $\mathrm{m}=-1$ radial and $\mathrm{m}=-2$ vertical disk-oscillation modes}.

In the past years, the RP model has been considered among the candidates for explaining the twin-peak QPOs in several LMXBs and related constraints on the sources have been discussed \citep[see, e.g.,][]{kar:1999,Zha-etal:2006:MONNR:kHzQPOFrCorr,bel-etal:2007a,Lam-Bou:2007:ASSL:ShrtPerBS,bar-bou:2008a,yan-etal:2009}. \change{While some of the early works discuss these constraints in terms of both NS mass and spin and include also the NS oblateness \citep[][]{mor-ste:1999,ste-etal:1999}, most of the published implications for individual sources focus on the NS mass and neglect its rotation.}

Two simultaneous kHz QPOs with centroid frequencies of up to 225 (500) Hz have recently been also found by \cite{bou-etal} in 11 different epochs of {{RXTE}}\,/\,Proportional Counter Array observations of the peculiar Z-source Circinus X-1.
Considering the RP model they reported the implied NS mass to be $M\sim2.2M_\odot$. The estimate was obtained assuming the non-rotating Schwarzschild spacetime and was based on fitting the observed correlation between the upper QPO frequency and the frequency difference $\Delta\nu=\nuU-\nuL$. In this paper, we improve the analysis of mass estimate carried out by Boutloukos et al. \change{In particular, we consider rotating spacetimes that comprehend the effects of frame-dragging and fit directly the correlation between the twin QPO frequencies. We show that good fits can be reached for the mass--angular-momentum relation rather than for the preferred combination of mass and spin.}

\section{Determination of Mass}

\change{Spacetimes around rotating NSs can be with a high precision approximated by the three parametric Hartle--Thorne (HT) solution of Einstein field equations (\citeauthor{har-tho:1968}, \citeyear{har-tho:1968}; see \citeauthor{ber-etal:2005}, \citeyear{ber-etal:2005}). The solution considers mass $M$, angular momentum $J$ and quadrupole moment $Q$ (supposed to reflect the rotationally induced oblateness of the star).  It is known that in most situations modeled with the present NS equations of state (EoS) the NS external geometry is very different from the Kerr geometry (representing the \qm{limit} of HT geometry for $\tilde{q}\equiv QM/J^2\rightarrow1$). However, the situation changes when the NS mass approaches maximum for a given EoS. For high masses the quadrupole moment does not induce large differences from the Kerr geometry since $\tilde q$ takes values close to unity (Appendix~\ref{appendix:Kerr:elaborate}).}

\change{The previous application of the RP model mostly implied rather very large masses \citep[e.g.][]{bel-etal:2007a}. These large masses are only marginally allowed by standard EoS. Also the mass inferred by \cite{bou-etal} takes values above $2M_\odot$. Motivated by this we use the limit of two-parametric Kerr geometry to estimate the influence of the spin of the central star in Circinus X-1 (see Appendix~\ref{appendix:Kerr:elaborate} where we pay a more detailed attention to rationalization and discussion of this choice allowing usage of simple and elegant Kerr formulae.}

\subsection{\change{Frequency Relations}}

\change{Assuming a compact object of mass  $M_{\mathrm{CGS}}=GM/c^2$ and dimensionless angular momentum  $j=cJ/GM^2$ described by the Kerr geometry, the explicit formulae for angular velocities related to Keplerian and radial frequencies are given by the following relations \citep[see][or \citeauthor{tor-stu:2005}, 2005]{ali-gal:1981, kat-etal:1998}
\begin{eqnarray}
\label{equation:Keplerian:radial}
\Omega_{\mathrm{K}} = \B(x^{3/2}+j)^{-1}\,,~\omega_{r}^2 = \Omega_{\mathrm{K}}^2\,\left(1-\frac{6}{x}+\frac{8j}{x^{3/2}}-\frac{3j^2}{x^2}\right)\,,
\end{eqnarray}
where $\B \equiv c^3/(2\pi GM)$ is the "relativistic factor" and $x\equiv r/M_\mathrm{CGS}$.
Considering Equations (\ref{equation:stella}) and (2), we can write for $\nuL$ and $\nuU$, both expressed in Hertz (see also Appendix~\ref{section:formulae} where we discuss a linear expansion of this formula),} 
\begin{eqnarray}
\label{equation:Kerr}
\nuL = \nuU\left\{1 - \left[1 + \frac{8j\nuU}{\B - j\nuU} - 6\left(\frac{\nuU}{\B - j\nuU}\right)^{2/3}\right.\right. \nonumber\\
           \left.\left. - 3j^2\left(\frac{\nuU}{\B - j\nuU}\right)^{4/3} \right]^{1/2}\right\}\,.\quad\quad\quad\quad\quad
\end{eqnarray}
In the Schwarzschild geometry, where $j=0$, Equation (\ref{equation:Kerr}) simplifies to
\begin{equation}
\label{equation:schwarszchild}
\nuL = \nuU\left\{1 - \left[1 - 6\left(\frac{\nuU}{\B}\right)^{2/3}\right]^{1/2}\right\}
\end{equation}
leading to the relation
\begin{equation}
\label{equation:difference}
\Delta\nu=\nuU\sqrt{1-6\left(2\pi GM\nuU\right)^{2/3}/c^2}
\end{equation}
that was used by Boutloukos et al. for the mass determination.\\\\

\begin{figure*}[t!]
\begin{minipage}{1\hsize}
\begin{center}
\hfill
\includegraphics[width=.415\textwidth]{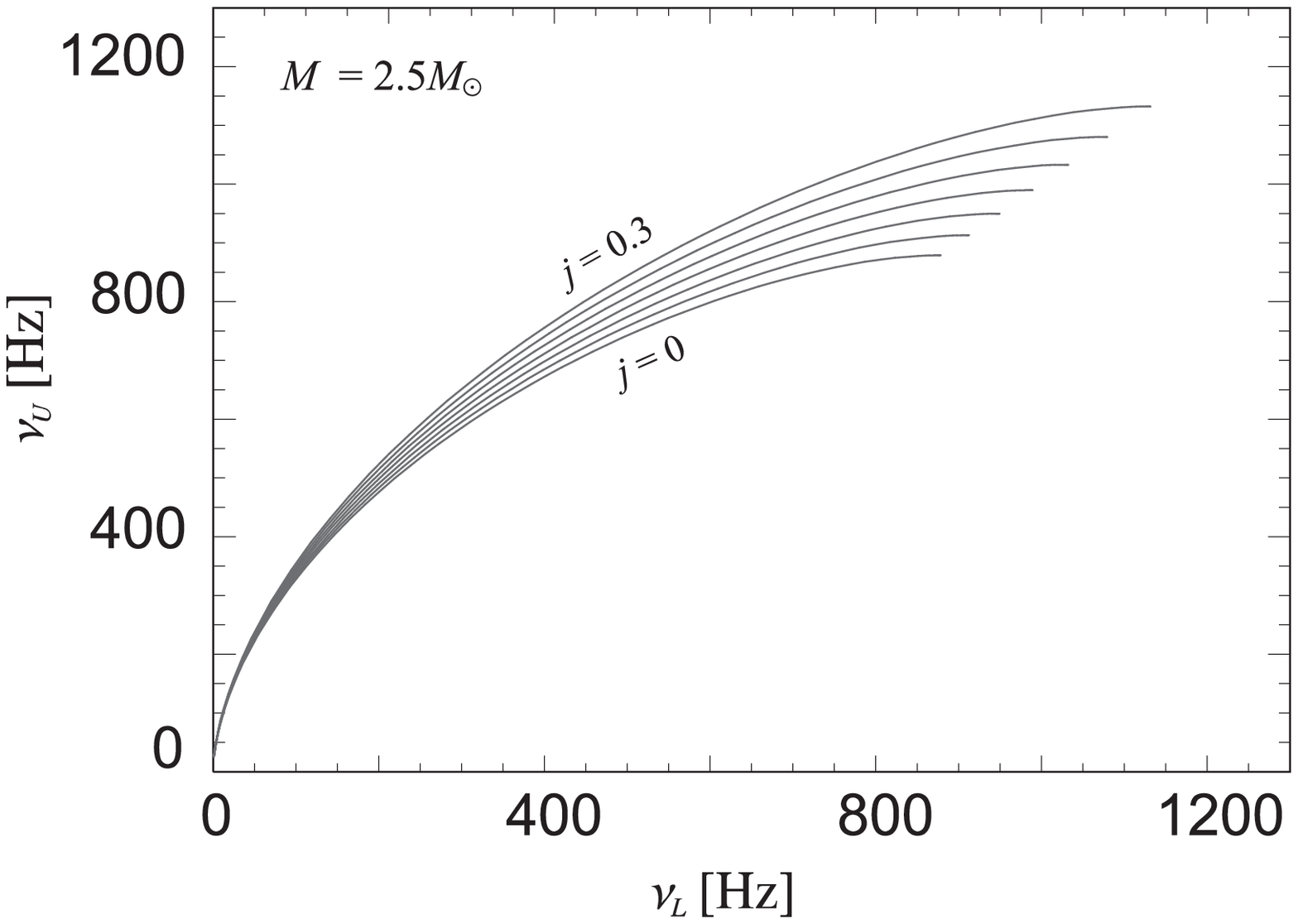}
~~~~
\includegraphics[width=.415\textwidth]{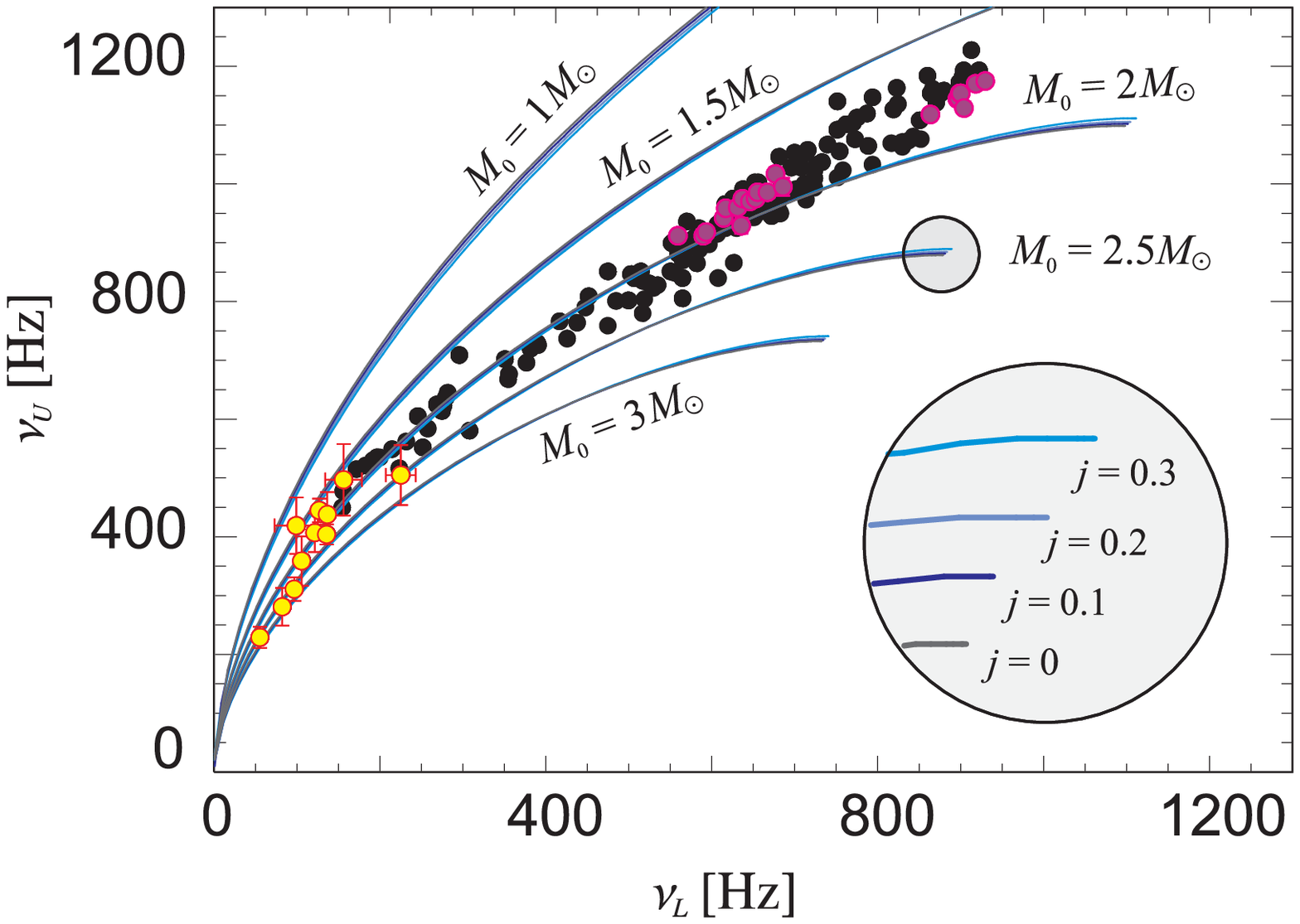}
\hfill~
\end{center}
\end{minipage}
\caption{ Left:  relation between the upper and lower QPO frequency following from the RP model for the mass $M=2.5M_\sun$. 
The consecutive curves differ in $j\in(0,\,0.3)$ by 0.05. Right: relations predicted by the RP model vs. data of several NS sources. The curves are plotted for various combinations of $M$ and $j$ given by  Equation (\ref{equation:quadratic}) with $k=0.7$. The datapoints belong to Circinus X-1 (red/yellow color), 4U 1636-53 (purple color) and most of other Z- and atoll- sources (black color) exhibiting large population of twin-peak QPOs.}
\label{figure:correlations}
\bigskip
\begin{minipage}{1\hsize}
\begin{center}
\includegraphics[width=0.85\textwidth]{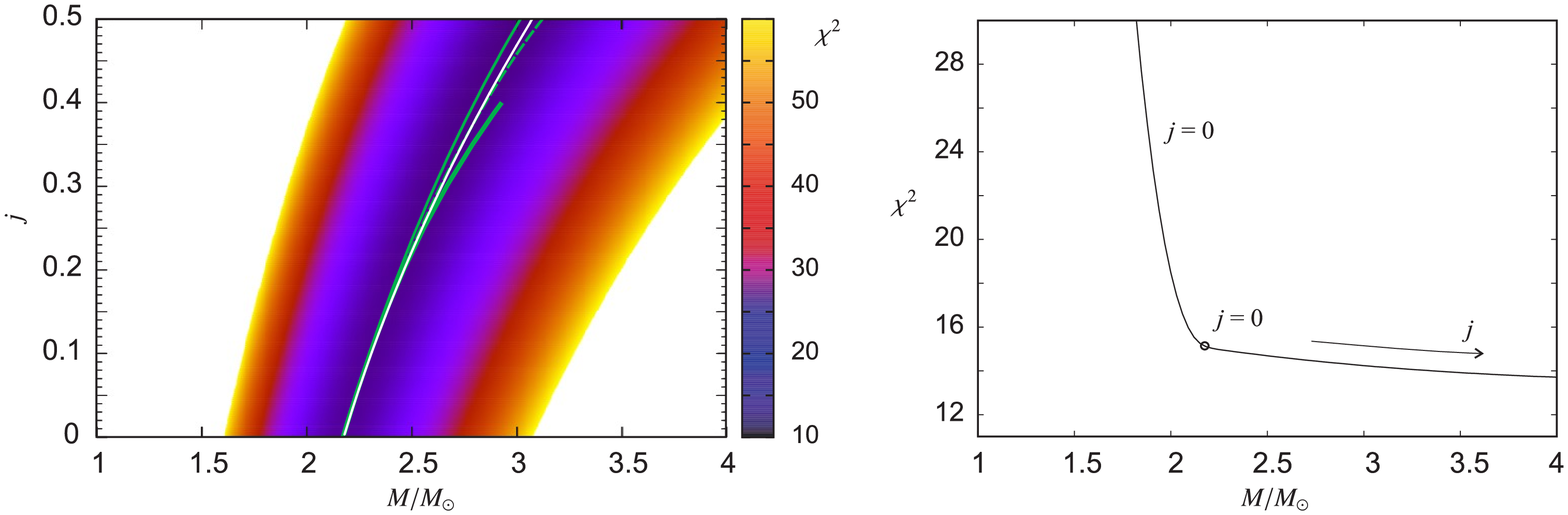}
\end{center}
\end{minipage}
\caption{ Left: \change{$\chi^2$ dependence on the parameters $M$ and $j$ assuming Kerr solution of Einstein field equations. The continuous white curve indicates the mass--angular-momentum relation (\ref{equation:mj:circ}). The continuous thin green curve denotes $j$ giving the best $\chi^2$ for a fixed $M$. The dashed and thick green curve indicates the same dependence but calculated using formulae (\ref{equation:Keplerian:radial:linear}) and (\ref{equation:linearized}) linear in $j$, respectively. The reasons restricting the calculation of the thick curve up to $j=0.4$ are discussed in Section \ref{section:formulae}.} Right: related profile of the best $\chi^2$ for a fixed $M$. The arrow indicates increasing  $j$.}  
\label{figure:map1}
\end{figure*}

\subsection{\qm{Ambiguity} in $M$}

There is a unique curve given by Equation~(\ref{equation:Kerr}) for each different combination of $M$ and $j$ (see Appendix \ref{appendix:ISCO} for the proof). The frequencies $\nuL$ and $\nuU$ scale as $1/M$ and, as illustrated in the left panel of Figure \ref{figure:correlations}, they increase with growing $j$. Naturally, one may ask an interesting question whether for different values of $M$ and $j$ there exist some curves that are similar to each other. We investigate and quantify this task in Appendix \ref{appendix:ISCO}.

There we infer\footnote{We first consider a special set of apparently similar curves sharing the terminal points. The set is (numerically) given by the particular choice of $M$, for any $j$ implying the same orbital frequency at the marginally stable circular orbit. The curves then only slightly differ in their concavity that increases with growing $j$.} 
that for $j$ up to $\sim0.3$ one gets a set of nearly identical \emph{integral curves} where $M$, $j$ and $M_0$ roughly relate as follows
\begin{equation}
\label{equation:quadratic}
M=[1+k(j+j^2)]\MS
\end{equation}
 with $$k=0.7\,.$$
 This result is illustrated in the right panel of Figure \ref{figure:correlations}. Clearly, when using relation (\ref{equation:quadratic}), any curve plotted for a rotating star of a certain mass can be well approximated by those plotted for a non-rotating star with a smaller mass, and vice versa.
Furthermore we find that (see Appendix \ref{appendix:ISCO}) when the \emph{top parts} of the curves (corresponding to $\nuU/\nuL\sim1\fromto1.5$ ) are considered only, the best similarity is reached for $$k=0.75\,.$$ 
These parts of the curves are potentially relevant to most of the atoll and high-frequency Z-sources data. On the other hand, for the (bottom) parts of the curves that are potentially relevant to low-frequency Z-sources including Circinus X-1, the best similarity is achieved for $$k=0.65\,(0.55,~0.5)\quad\mathrm{when}\quad\nuU/\nuL\sim 2\,(3,~4)\,.$$

{Taking into account the above consideration we can expect that the single-parameter best fit to the data by relation (\ref{equation:schwarszchild}) roughly determines a set of mass-angular-momentum combinations (\ref{equation:quadratic}) with similar $\chi^2$.} The result of Boutloukos et al. then implies that good fits to their data, displaying $\nuU/\nuL\sim 3$, should be reached for $M\sim2.2M_{\odot}[1+0.55(j+j^2)]$. In what follows we fit the data and check this expectation.

\newpage

\subsection{Data Matching}

In the right panel of Figure \ref{figure:correlations}, we show the twin-peak frequencies measured in the several atoll and Z-sources\footnote{After \citet{bar-etal:2005a,bar-etal:2005b, Boirin-00,bel-etal:2007a,diSalvo-03,Homan-02,Jonker2002a,Jonker2002b,MvdK2000,M2001,van-Straaten-00,van-Straaten-02,Zhang-98}.} together with the observations of Circinus X-1. For the Circinus X-1 data we search for the best fit of the one-parametric relation (\ref{equation:schwarszchild}). Already from Figure \ref{figure:correlations}, where these data are emphasized by the red/yellow points, one may estimate that the best fit should arise for $M_0\in2\fromto2.5M_{\sun}$. Using the standard least squares method \citep[][]{pre-etal:2007} we find the lowest $\chi^2\doteq15\doteq2\,$dof for the mass $M_0\,\doteq\,2.2M_{\sun}$ which is consistent with the value reported by Boutloukos et al. The \change{symmetrized} error corresponding to the unit variation of $\chi^2$ is $\pm0.3M_\sun$. \change{The asymmetric evaluation of $M_0$ reads $2.2[+0.3;-0.1]M_{\sun}$.} The \change{white} curve in Figure \ref{figure:map1} indicates the mass-angular-momentum relation implied by Equation (\ref{equation:quadratic}),
\begin{equation}
\label{equation:mj:circ}
M=2.2M_{\odot}[1+k(j+j^2)],\quad k=0.55.
\end{equation}

For the exact fits in Kerr spacetime we calculate the relevant frequency relations for the range of $M\in\!1$--$4M_{\sun}$ and $j\in\!0$--$0.5$. These relations are compared to the data in order to calculate the map of $\chi^2$. We use the step equivalent to thousand points in both parameters and obtain a two-dimensional map of $10^6$ points. This color-coded map is included in the left panel of Figure~\ref{figure:map1}. One can see in the map that the acceptable $\chi^2$ is rather broadly distributed. The \change{thin solid} green curve indicates $j$ corresponding to the best $\chi^2$ for a fixed $M$. It well agrees with the expected relation (\ref{equation:mj:circ}) denoted by the \change{white} curve. The right panel of Figure~\ref{figure:map1} then shows in detail the dependence of the best $\chi^2$ for the fixed $M$. It is well visible that the quality of the fit tends to very gently, monotonically increase with rising  $j$  and it is roughly  $\chi^2\sim15$ for any considered $j$.
\smallskip

\section{\change{Discussion and Conclusions}}
\label{section:conclusions}

\change{The quality of the fit tends to very gently, monotonically increase with rising $j$  and it is roughly
\begin{equation}
\label{equation:mj:final}
 \chi^2\sim2\,\mathrm{dof}\Leftrightarrow M\sim2.2[+0.3,-0.1]M_{\odot}\times[1+0.55(j+j^2)].
\end{equation} 
{Therefore one cannot estimate the mass without the independent knowledge of the spin or vice versa, and the above relation provides the only related information implied by the geodesic RP model.}} 

\change{To obtain relation (\ref{equation:mj:final}), the exact Kerr solution of Einstein field equations was considered. The choice of this two-parametric spacetime description and related formulae (\ref{equation:Keplerian:radial}) is justified by a large value of the expected mass $M_0$ (see Appendix~\ref{appendix:Kerr:elaborate} for details). In Appendix~\ref{section:formulae} we discuss the utilization of the linearized frame-dragging description. Figure~\ref{figure:map1} includes the mass--spin dependence giving best $\chi^2$ resulting when the fitting of datapoints is based on the associated formulae (\ref{equation:Keplerian:radial:linear}) respectively (\ref{equation:linearized}).  Considering that $\nuL(\nuU)$ formula (\ref{equation:Kerr}) merge up to the first order in $j$ with the $\nuL(\nuU)$ relation (\ref{equation:linearized}) linear in $j$ one can expect that the associated $M(j)$ relations obtained from fitting of data should roughly coincide up to $j\sim0.1\fromto0.2$. From the figure we can find that there is not a big difference between the resulting $M(j)$ relations even up to much higher $j$. The extended coincidence can be clearly explained in terms of the kHz QPO frequency ratio $R\equiv\nuU/\nuL$.\footnote{\change{Orbital frequencies scale with $1/M$. For any model considering $\nuL$ and $\nuU$ given by their certain combination, the ratio $R$ represents the measure of radial position of the QPO excitation (provided that the NS spin and EoS are fixed).}}} 

\change{Observations of Circinus X-1 result to $R\sim2.5\fromto4.5$ while usually it is $R\sim1.2\fromto3$ \citep[and most often $R\sim1.5$, ][]{abr-etal:2003b,tor-etal:2008b,yan-etal:2009}. Assuming the RP model along with any $j\in(0,~1)$, the ratio $R=2$ corresponds with a good accuracy to radii where the radial epicyclic frequency reaches its maximum  \citep{tor-etal:2008a}.  Only values lower than $R\sim2$ are then associated with the proximity of innermost stable circular orbit (ISCO) where the effects of frame dragging become to be highly non-linear in both $j$ and $r$. Accordingly, for a given $j$, in the case when $R\sim3$, the individual formulae restricted up to certain orders in $j$ are already close to their common linear expansion in $j$ and differ much less than for $R\sim1.5$ (see Appendix~\ref{appendix:Kerr:elaborate}).}

\change{The rarely large $R$ and associated high radial distance \citep[both already remarked by][although in a different context]{bou-etal} in addition to large $M_0$ warrant relevance of relation (\ref{equation:mj:final}) for rather high values of the angular momentum. {Consequently, we can firmly conclude that the upper constrained limit of the mass changes from the value $2.5M_{\sun}$ to $3M_{\sun}$ for $j=0.3$ and even to $3.5M_{\sun}$ for $j=0.5$.} The Value of $M_0$ that is above $2M_{\sun}$ and the increase of $M$ with growing $j$ for corotating orbits elaborated here are challenging for the adopted physical model. Further detailed investigation involving realistic calculations of the NS structure can be therefore effective in relation to EoS selection or even falsifying the RP model.}

{{Finally, we note that discussed trend of increase of estimated mass arising due to rotational effects should be relevant also for several other sources.} Of course, many systems display mostly low values of $R$. These low values of $R$ are in context of the RP model suggestive of proximity of ISCO. \citet{tor:2009} and \citet{zha-etal:2010} pointed that under the consideration of the RP model and $j=0$, most of the high-frequency sources data are associated with radii close to $r=6.75M$. Possible signature of ISCO in high frequency sources data has been also reported in a series of works by \citeauthor{bar-etal:2005a} (\citeyear{bar-etal:2005a,bar-etal:2005b,bar-etal:2006}) based on a sharp drop in the frequency behavior of the kHz QPO quality factors \citep[for instance the atoll source 4U 1636-53 denoted by \qm{blueberry} points in Figure \ref{figure:correlations}) clearly exhibits both low $R$ and a drop of QPO coherence, see][]{bou-etal:2009}. Considering the proximity of ISCO, high-order non-linearities in both $j$ and $r$ are important and even small differences between the actual NS and Kerr metric could have certain relevance. For this reason some caution is needed when applying our results to high frequency sources.}

\acknowledgments
This work has been supported by the Czech grants MSM~4781305903, LC~06014, and GA\v{C}R~202/09/0772. The authors thank to the anonymous referee for his objections and comments which helped to greatly improve the paper. We also appreciate useful discussions with Milan \v{S}enk\'{y}\v{r}.


\appendix               

\section{A. Approximations, Formulae and Expectations}

\subsection{A.1.~\change{Matching Influence of Neutron Star Spin}}
\label{appendix:Kerr:elaborate}
 
\change{
Rotation and the related frame-dragging effects strongly influence the processes in the vicinity of compact objects and there is a need of their reflection in the appropriate spacetime description. External metric coefficients related to up-to-date sophisticated  models of rotating NS are taken out of the model in two distinct ways. In the first way, the coefficients are obtained \qm{directly} from differential equations solved inside the numeric NS model, while in the second (more usual) way they are inferred from the main parameters of the numeric model (mass, angular momentum etc.) through an approximative analytic prescription.
Several commonly used numerical codes related to rotating NS have been developed and discussed (see, RNS: \citeauthor{RNS}, \citeyear{RNS}; LORENE: \citeauthor{LORENE}, \citeyear{LORENE}; and also \citeauthor{noz-etal:1998}, \citeyear{noz-etal:1998}; \citeauthor{ste-fri:1995}, \citeyear{ste-fri:1995}; \citeauthor{coo-etal:1994}, \citeyear{coo-etal:1994}; \citeauthor{kom-etal:1989}, \citeyear{kom-etal:1989}).}

\subsubsection{\change{A.1.1. Analytical Approximations and High-mass Neutron Stars}}

\change{
In the context of a simplified analysis of NS frame-dragging consequences, an approximation through two solutions of Einstein field equations is usually recalled: \emph{Lense-Thirring metric} also named linear-Hartle metric \citep[][]{thi-len:1918,har-sha:1967,har:1967} and \emph{Kerr-black-hole metric} together with related formulae \citep{ker:1963,boy-lin:1967,car:1971,bar-etal:1972}. It is expected that the Lense-Thirring metric well fits the most important changes (compared to the static case) in the external spacetime structure of a \emph{slowly rotating NS}. This expectation is usually assumed for $j<0.1\fromto0.2$.\footnote{\change{The interval $0<j<2\times10^{-1}$ is often assumed as one of the several possible definitions of \qm{\emph{slow rotation}}. However, in relation to implications of the frame-dragging effects, the effective size of this interval depends on the radial coordinate. For $x$ close or below $x_\mathrm{ms}$ the interval in $j$ rather reduces to low values. On the other hand for $x$ above the radius of the maximum of $\nu_\mathrm{r}$ the interval can be extended to $j$ higher than $j=0.2$. The term slow rotation is also frequently considered in another context. For instance when using the HT metric in NS models the slow rotation is usually associated with the applicability of the metric and consequently to spins up to $\sim\!800$Hz for most of EOS and NS masses. For these reasons we do not use the term elsewhere in the paper.}} Due to asymptotical flatness constraints the formulae related to Lense-Thirring, Kerr and some other solutions considered for rotating NS merge when truncated to the first order in $j$. Accordingly, for the astrophysical purposes there is a widespread usage of the approximative terms derived with the accuracy of the first order in $j$.}
\change{While these approximations are two-parametric, the more realistic approximations; for instance those given by the HT metric \citep[][]{har-tho:1968} and related terms \citep{abr-etal:2003a}, relations of \cite{shi-sas:1998} or the solution of \cite{pac-etal:2006}, deal with more parameters and provide less-straightforward formulae. Perhaps also because of that they are not often considered in discussions of concrete astrophysical compact objects.}

\begin{figure*}[t!]
\begin{minipage}{1\hsize}
\begin{center}
~~~~~~~~~\includegraphics[width=.84\textwidth]{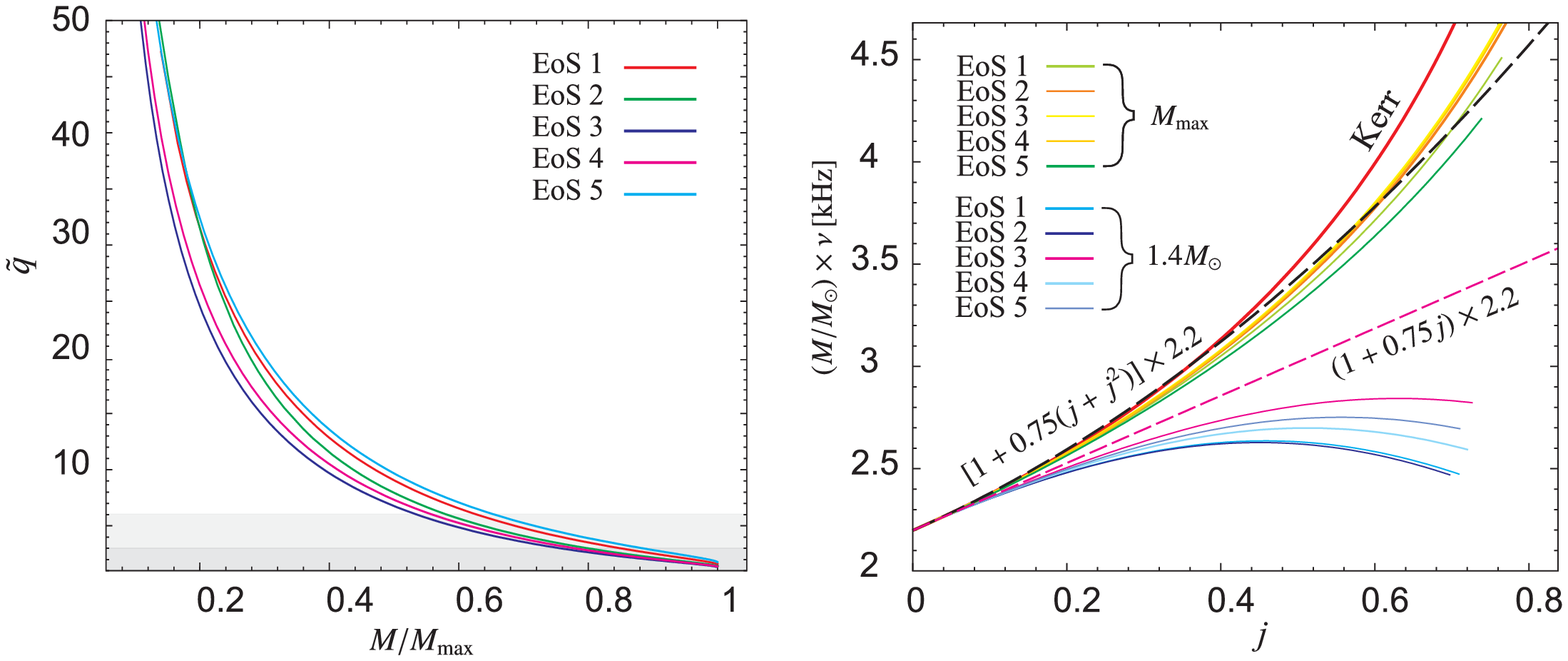}
\end{center}
\end{minipage}
\caption{\change{Left: parameter $\tilde{q}$ for several EoS. Shaded areas denote $\tilde{q}=6$ and $\tilde{q}=3$. \mbox{---~Right:} ISCO frequencies for the same EoS as used in the left panel. The curves are calculated for mass $1.4M_\odot$ and a relevant maximal allowed mass. The curves following from the exact Kerr solution and linear relation (\ref{equation:ISCO:kluzniak}) are displayed as well. The quadratic relation denoted by the black-dashed curve is discussed later in Section \ref{appendix:ISCO:formulae}.}}
\label{figure:EOS}
\bigskip
\begin{minipage}{1\hsize}
\begin{center}
\hfill~
\includegraphics[width=.4\textwidth]{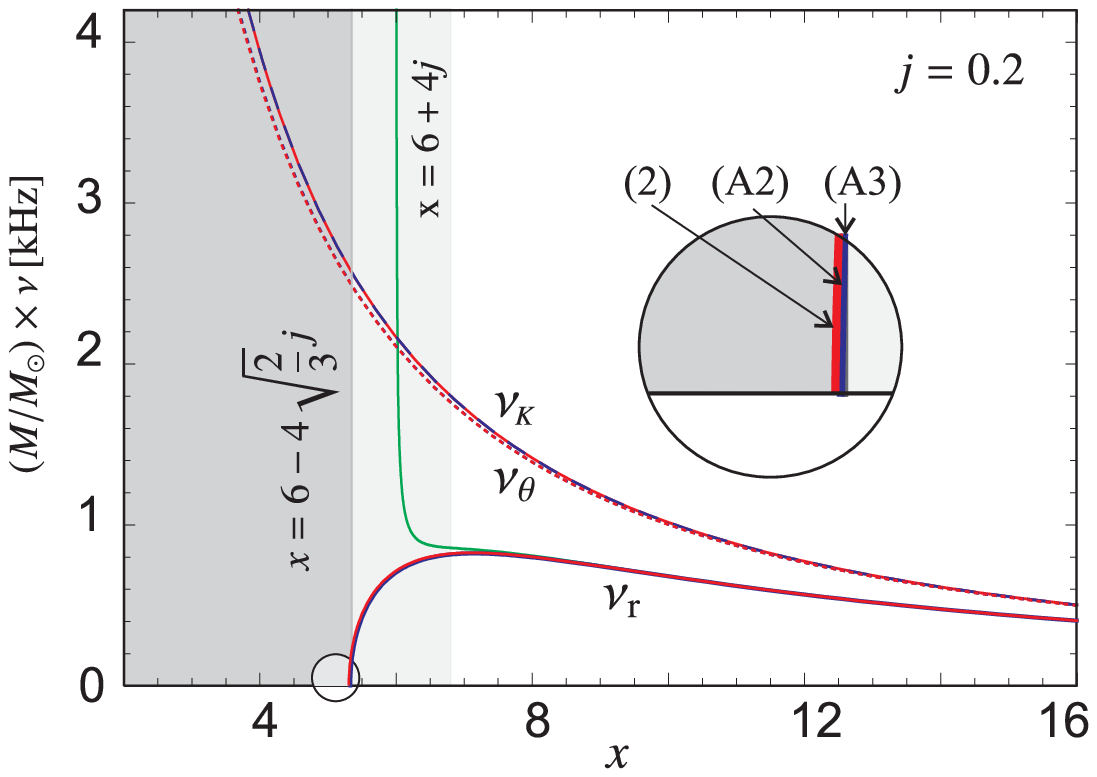}
\hfill
\includegraphics[width=.4\textwidth]{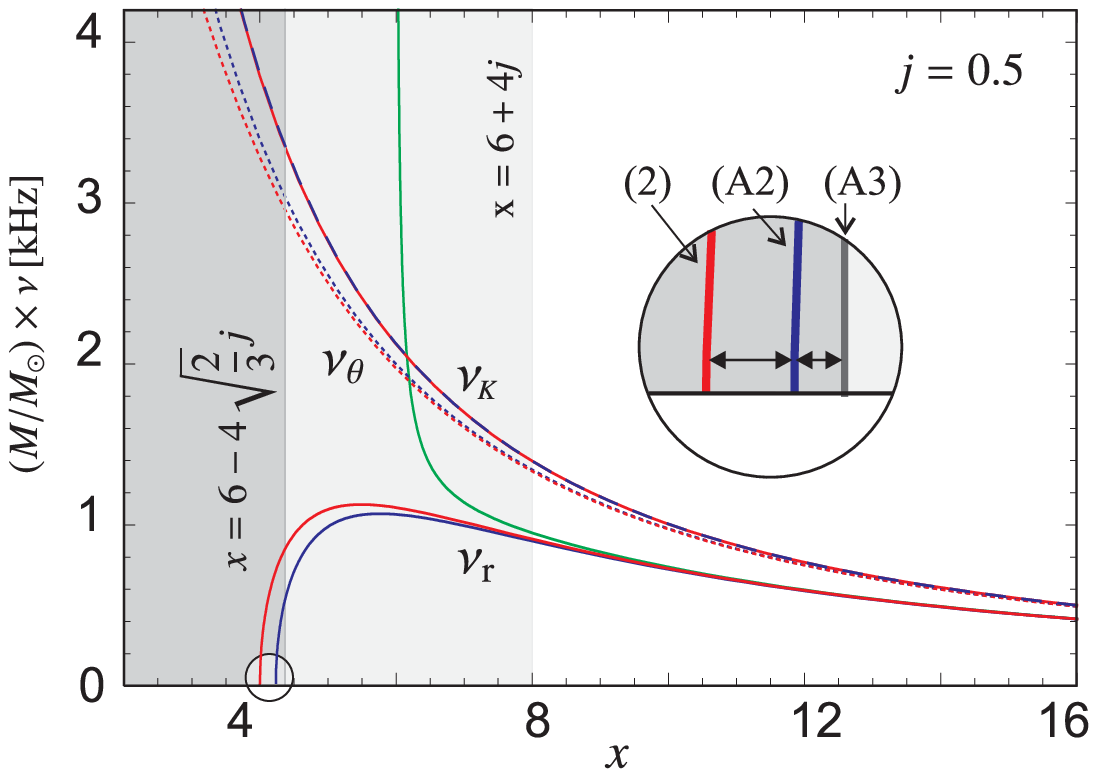}
\hfill~
\end{center}
\end{minipage}
\caption{\change{Frequencies of the perturbed circular geodesic motion. Relations for the Kerr metric given by Equation (\ref{equation:Keplerian:radial:linear}) are denoted by blue and dashed-blue curves. Relations (\ref{equation:Keplerian:radial:linear}) are indicated by red curves while relation (\ref{equation:nurl}) is plotted using the green color. Dotted relations denote the Kerr- and linearized- vertical frequencies that are not discussed here \citep[see][]{mor-ste:1999,ste-etal:1999}. Inset emphasizes a difference between the radii fulfilling the ISCO condition $\nur=0$ for the relations (\ref{equation:Keplerian:radial}) explicitly given by Equation (\ref{equation:rms}), Equation (\ref{equation:Keplerian:radial:linear}) and the ISCO-radius given by Equation (\ref{equation:rms:linear}).}}
\label{figure:epicyclic}
\bigskip
\begin{minipage}{1\hsize}
\begin{center}
\hfill~
\includegraphics[width=0.4\textwidth]{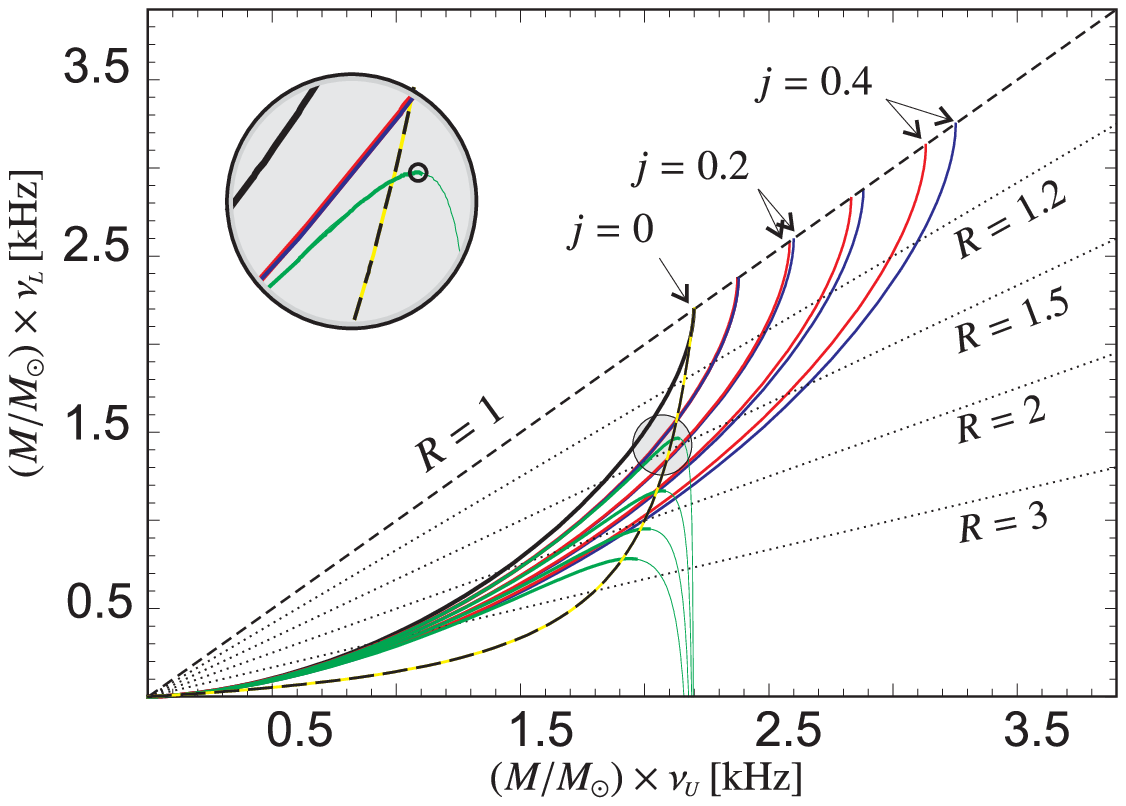}
\hfill
\includegraphics[width=.4\textwidth]{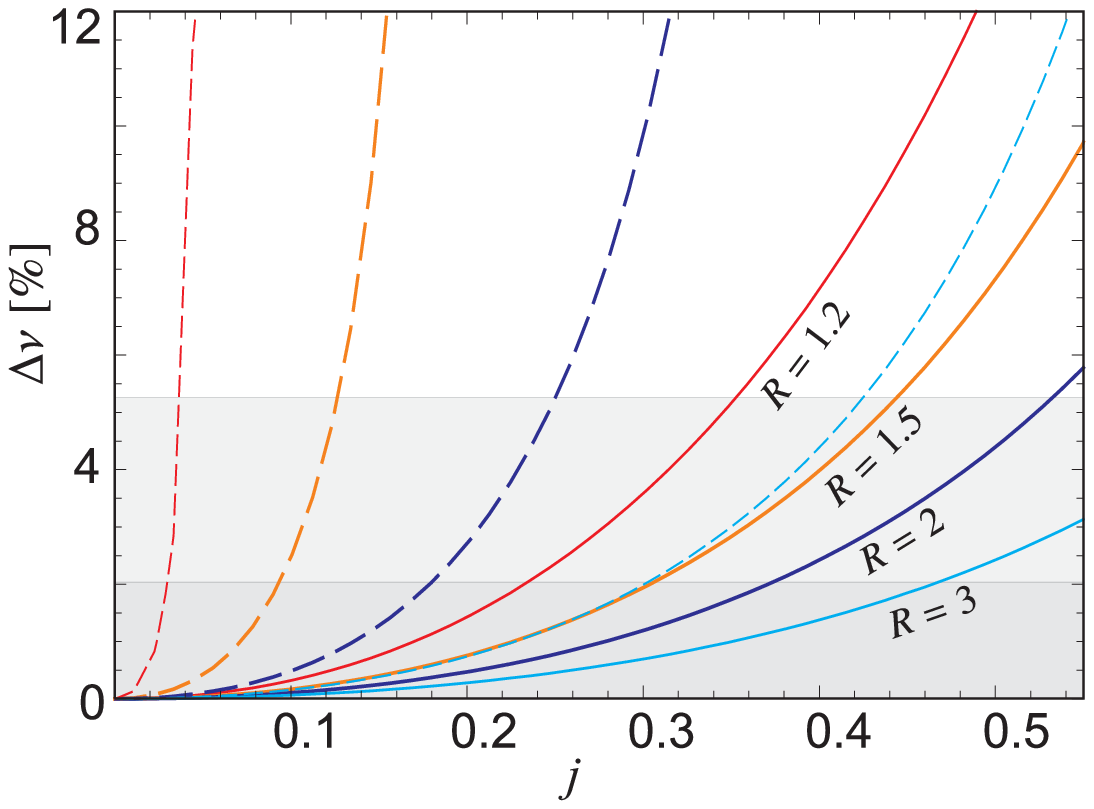}
\hfill~
\end{center}
\end{minipage}
\caption{\change{Left: the RP model frequency relations given by Equation (\ref{equation:Kerr}), blue curves; formulae (\ref{equation:Keplerian:radial:linear}), red curves; relations (\ref{equation:linearized}), green curves. Relation (\ref{equation:condition}) roughly determining the applicability of (\ref{equation:linearized}) is denoted by the dashed black/yellow curve. Right: related differences $\Delta \nu$ between the lower QPO frequency implied by the Kerr formulae (\ref{equation:Kerr}) and those following from Equation (\ref{equation:Keplerian:radial:linear}) respectively Equation (\ref{equation:linearized}) indicated by continuous respectively dashed curves. Different colors correspond to different frequency ratio $R$. Shaded areas indicate $\Delta\nu<5\%$ and $\Delta\nu<2\%$.}}
\label{figure:relations}
\end{figure*}

\change{
Astrophysical applicability of the above analytical approaches has been extensively tested in the past ten years. Criteria based on the comparison of miscellaneous useful quantities have been considered for these tests \citep[e.g.,][]{mil-etal:1998b,ber-etal:2005}. It has been found that  spacetimes induced by most of up-to-date NS EoS without inclusion of magnetic field effects are well approximated with the HT solution of the Einstein field equations \citep[see][for details]{ber-etal:2005}. The solution reflects three parameters: NS mass $M$, angular momentum $J$ and quadrupole moment $Q$.  Note that Kerr geometry represents the \qm{limit} of the HT geometry for $\tilde{q}\equiv Q/J^2\rightarrow1$. The parameter $\tilde{q}$ then can be used to characterize the diversity between the NS and Kerr metric.}

\change{
The left panel of Figure \ref{figure:EOS} displays a dependence of $\tilde{q}$ on the NS mass. This illustrative figure was calculated following  \citet{har:1967}, \citet{har-tho:1968}, \citet{cha-mil:1974} and \citet{mil:1977}. The considered EoS are denoted as follows (see \citeauthor{lat-pra:2001}, \citeyear{lat-pra:2001} and \citeauthor{lat-pra:2007}, \citeyear{lat-pra:2007}  for details):}
\\\change{[EoS1]} \change{SLy 4, \cite{rik-etal:2003}.}
\\\change{[EoS2]} \change{APR, \cite{akm-etal:1998}.}
\\\change{[EoS3]} \change{AU (WFF1), \cite{wir-etal:1988,ste-fri:1995}.}
\\\change{[EoS4]} \change{UU (WFF2), \cite{wir-etal:1988,ste-fri:1995}.}
\\\change{[EoS5]} \change{WS (WFF3), \cite{wir-etal:1988,ste-fri:1995}.}

\change{
Inspecting the left panel of Figure \ref{figure:EOS} we can see that for EoS configurations  resulting to low- or middle-mass of the central star ($M$ up to 0.8$M_\mathrm{max}$, i.e., roughly up to $1.4M\fromto 1.8M_{\sun}$) depending on EOS, the implied HT geometry is rather different from the Kerr geometry. More specifically, for a fixed central density, $\tilde{q}$ strongly depends on the given EoS and substantially differs from unity. On the contrary, for \emph{high mass configurations} $\tilde{q}$ approaches unity implying that the actual NS geometry is close to Kerr geometry. One can expect that in such case formulae related to the Kerr geometry should provide better approximation than for low values of $M$. Next, focusing on high-mass NS, we briefly elaborate some points connected to the applicability of the Kerr formulae and related linearized terms.}

\subsubsection{\change{A.1.2. Kerr and linearized Kerr Formulae: Comparison, Utilization and Restrictions}}
\label{section:formulae}

The radial epicyclic frequency  goes to zero on a particular, so-called \emph{marginally stable circular orbit} $x_\mathrm{ms}$ \citep[e.g.,][]{bar-etal:1972}. \change{In Kerr spacetimes} it is given by the relation \citep[][]{bar-etal:1972}
\begin{eqnarray}
\label{equation:rms}
&&x_\mathrm{ms} = 3+Z_{\,2}-\sqrt{(3-Z_{\,1})(3+Z_{\,1}+2Z_{\,2})}\,,\\
&&\mathrm{where}\nonumber\\
&&Z_{\,1} = 1+(1-j^2)^{1/3}\left[(1+j)^{1/3}+(1-j)^{1/3}\right]\,,\nonumber\\
&&Z_{\,2} = \sqrt{3j^2+Z_{\,1}^2}\,.\nonumber
\end{eqnarray}
\change{Below $x_{\mathrm{ms}}$ there is no circular geodesic motion stable with respect to radial perturbations.} The orbit is often named ISCO and determines the inner edge of a thin accretion disk. The corresponding ISCO orbital frequency $\nuK(x_\mathrm{ms})$ represents the highest possible orbital frequency of the thin disk and the related \qm{spiraling} inhomogenities \citep{klu-etal:1990}. 
\change{Dependence of ISCO frequency on $j$ following from Equation (\ref{equation:rms}) is shown in the right panel of Figure \ref{figure:EOS}.}

\change{Assuming the description of geodesic motion accurate in the first order of $j$, using Taylor expansion around $j=0$, one may rewrite the explicit terms in Equation (\ref{equation:Keplerian:radial}) as 
\begin{eqnarray}
\label{equation:Keplerian:radial:linear}
\Omega_{\mathrm{K}}=\B\left(\frac{1}{x^{3/2}} -\frac{j}{x^3}\right)\,, \quad
\omega_{r}^2 = \Omega_{\mathrm{K}}^2\,\left(1-\frac{6}{x}+\frac{8j}{x^{3/2}}\right)\,.
\end{eqnarray}
Consequently, \emph{linearized} formula for the ISCO radius can be expressed as
\begin{equation}
\label{equation:rms:linear}
x_\mathrm{ms} = 6-4 \sqrt{\frac{2}{3}} j\,.
\end{equation}
Note that the root of the expression for $\omega_\mathrm{r}^2$  from Equation (\ref{equation:Keplerian:radial:linear})  is of higher order in $j$ so that the exact radius where $\omega_\mathrm{r}$ vanishes  agrees with the solution (\ref{equation:rms:linear}) only in the first order of $j$. The related ISCO frequency can be evaluated as \citep[][]{klu-wag:1985,klu-etal:1990}
\begin{eqnarray}
\label{equation:ISCO:kluzniak}
\nuK(x_\mathrm{ms})&=&(M_\sun/M)\times(1+0.749j)\times\nuK(x_\mathrm{ms},\,M=M_\sun,\,j=0)
\nonumber\\
&\doteq&(M_\sun/M)\times(1+0.749j)\times2197\,\mathrm{Hz}.
\end{eqnarray}
This frequently considered relation is included in the right panel of Figure \ref{figure:EOS}.}

\change{
In the right panel of Figure \ref{figure:EOS} we integrate the ISCO frequencies plotted for several EoS (the same as in the left panel). We choose two groups of models - one calculated for the set of five different EoS and  \qm{canonic} mass $1.4M_\sun$, the other one for the same set of EoS but considering a maximal mass allowed by each individual EoS. This choice allows for the illustration of middle- and high- mass behavior of ISCO relations and  comparison of  their simple approximations. Clearly, for middle mass configurations  Equation (\ref{equation:ISCO:kluzniak}) provides  better approximation than using the Kerr-spacetime formulae \citep[see also][]{mil-etal:1998b}. On the other hand, when the high-mass configurations are considered, the Kerr solution provides better approximation than Equation (\ref{equation:ISCO:kluzniak}). Moreover, its accuracy is higher than the accuracy of both approximations for middle-mass configurations.}

\subsubsection{\change{A.1.3. Geodesic frequencies and RP model}}
\label{appendix:Kerr:RP}

\change{When the expression for the radial epicyclic frequency given by Equation (\ref{equation:Keplerian:radial}) or (\ref{equation:Keplerian:radial:linear}) is fully linearized in $j$, it leads to
\begin{equation}
\label{equation:nurl} 
\omega_\mathrm{r}=\frac{(x-6)x^{3/2}+3j(x+2)}{\sqrt{(x-6)x^7}}\,.
\end{equation}
Relation (\ref{equation:nurl}) provides a good approximation except for the vicinity of $x_\mathrm{ms}(j)$ as it diverges at $x=6$. Note that this divergence arises only for corotating but not counterrotating orbits (which we however do not discuss in this paper). For any positive $j<0.5$ the fully linearized frequency (Equation (\ref{equation:nurl})) does not differ from $\omega_r$ given by Equation (\ref{equation:Keplerian:radial:linear}) for more than about $5\%$ when $x\gtrsim 6+4j$. The left panel of Figure \ref{figure:epicyclic} compares the frequencies of geodesic motion associated directly with Kerr metric, to formulae (\ref{equation:Keplerian:radial:linear}) and (\ref{equation:linearized}), respectively.}

\change{{Assuming linearized Keplerian frequency given by Equation (\ref{equation:Keplerian:radial:linear}) and the radial epicyclic frequency (Equation (\ref{equation:nurl})), we can write for the RP model the relation between $\nuL$ and $\nuU$  as
\begin{equation}
\label{equation:linearized}
\nuL = \nuU \left(1-\sqrt{1-6 \alpha}+\frac{2 j \nuU \left(\alpha-2\right)}{\B \sqrt{1-6 \alpha}}\right),\quad\alpha=\left(\frac{\nuU}{\B}\right)^{2/3}\,,
\end{equation}
which equals the first-order expansion of Kerr spacetime Equation (\ref{equation:Kerr}) and also to the first-order expansion of the same relation if it would be derived for Lense--Thirring or HT metric.} 
Similarly to relation (\ref{equation:nurl}), relation (\ref{equation:linearized}) loses its physical meaning for frequencies close to $\nuK(\mathrm{ISCO})$ since it reaches a maximum at frequencies that can be expressed with a small inaccuracy as 
\begin{equation}
\label{equation:condition}
\nuL=\frac{\nuU}{12-{\nuU}/{200}}\left(\frac{M_{\sun}}{M}\right)\,(\mathrm{Hz})\,.
\end{equation}
The left panel of Figure \ref{figure:relations} compares the frequency relations (\ref{equation:linearized})  to  relations (\ref{equation:Kerr}) and those following from formulae (\ref{equation:Keplerian:radial:linear}). {It is useful to discuss their differences in terms of the frequency ratio $R=\nuU/\nuL$.} For a fixed $j$ the frequencies $\nuL$ and $\nuU$ scale with $1/M$. The ratio $R$ then represents a \qm{measure} of the radial position of the QPO excitation. It always reaches $R=1$ at ISCO where the non-linear $j$ terms are important and  $R=0$ at infinity where the spacetime is flat. Note that $R=2$ almost exactly corresponds to the maximum of $\nur$ for any $j$ \citep[][]{tor-etal:2008a}.} 

\change{The right panel of Figure \ref{figure:relations} quantifies differences between the QPO frequency implied by the Kerr formulae (\ref{equation:Kerr}), relations (\ref{equation:Keplerian:radial:linear}) and relation (\ref{equation:linearized}). We can see that differences between the Kerr relations (\ref{equation:Kerr}) and those implied by formulae (\ref{equation:Keplerian:radial:linear}) become small when $R\gtrsim 2$ ($\Delta\nu\lesssim 5\%$ for $j\lesssim0.5$).
{For $R\sim3$ and higher, relations (\ref{equation:Kerr}) and those implied by Equation (\ref{equation:Keplerian:radial:linear}) are almost equivalent nearly merging to their common linear expansion (\ref{equation:nurl}).} Note that  taking into account relation (\ref{equation:condition}) the linear expansion (Equation (\ref{equation:linearized})) provides reasonable physical approximation for spins and frequency ratios roughly related as 
\begin{equation}
\label{equation:condition:R}
j\lesssim 0.3(R-1).
\end{equation}
}

\subsubsection{\change{A.1.4. Applications}}

\change{Several values of NS mass previously reported to be required by the RP model, including the estimate of Boutloukos et al., belong to the upper part of the interval allowed by standard EoS. We can therefore expect low $\tilde{q}$ and take the advantage of the exact Kerr solution for most of the practical calculations needed through the paper. Unlike to formulae truncated to certain order, all the formulae derived from the exact Kerr solution are from the mathematical point of view fully selfconsistent for any $j$. This allows us to present the content of Appendix~\ref{appendix:ISCO} in a compact and demonstrable form. 
}

\change{In Section \ref{section:conclusions} we finally compare the results of QPO frequency relation fits for Circinus X-1 using the Kerr solution and those done assuming Equations (\ref{equation:Keplerian:radial:linear}) respectively (\ref{equation:linearized}). From the previous discussion it can be expected that for Circinus X-1, due to its exceptionally high $R$, the fits obtained with the Kerr formulae (\ref{equation:Kerr}) and \qm{linear} formulae (\ref{equation:Keplerian:radial:linear}) should nearly merge with the fits obtained assuming the common linear expansion (Equation (\ref{equation:linearized})). Note also that, on a technical side, the linear expansion can be used up to $j\sim0.3\fromto0.4$ since the lowest $R$ in the Circinus X-1 data is $R\sim2\fromto2.5$ (Equation (\ref{equation:condition:R})).}

\subsection{A.2. Uniqueness of Predicted Curves and \qm{Ambiguity} in $M$}
\label{appendix:ISCO}

\change{The radial epicyclic frequency vanishes at $x_{\mathrm{ms}}$. In the RP model it is then $\nuU^{\,\mathrm{max}} = \nuL^{\,\mathrm{max}} = \nuK(x_\mathrm{ms})$.} Obviously, if there are two different combinations of $M$ and $j$ which, based on the RP model, imply the same curve $\nuU(\nuL)$, such combinations must also imply the same ISCO frequency.

In the left panel of Figure \ref{figure:ISCO1} we show a set of curves constructed as follows. We choose \mbox{$M_0^*=2.5M_\sun$} and $j\in(0,\,0.5)$ and for each different $j$ we numerically find $M$ such that the corresponding ISCO frequency is equal to those for $M_0^*$ and $j=0$. Then we plot the $\nuU(\nuL)$ curve for each combination of $M$ and $j$. We can see that except for the terminal points the curves split. The frequencies in the figure can be rescaled for any \qm{Schwarzschild} mass $M_0$ as $M_0^*/M_0$. Thus, the scatter between the curves provides the proof that one cannot obtain the same curve for two different combinations of $M$ and $j$.

On the other hand, the discussed scatter is apparently small and the curves differ only slightly in the concavity that grows with increasing $j$. This has an important consequence. {The curves are very similar with respect to the typical inaccuracy of the measured NS twin-peak data and there arises a possible mass--angular-momentum ambiguity in the process of fitting the datapoints.} Next, we derive a simple relation approximating this ambiguity. 

\subsubsection{A.2.1. Formulae for ISCO Frequency}
\label{appendix:ISCO:formulae}

\begin{figure*}[t!]
\begin{minipage}{1\hsize}
\begin{center}
\hfill~
\includegraphics[width=.42\textwidth]{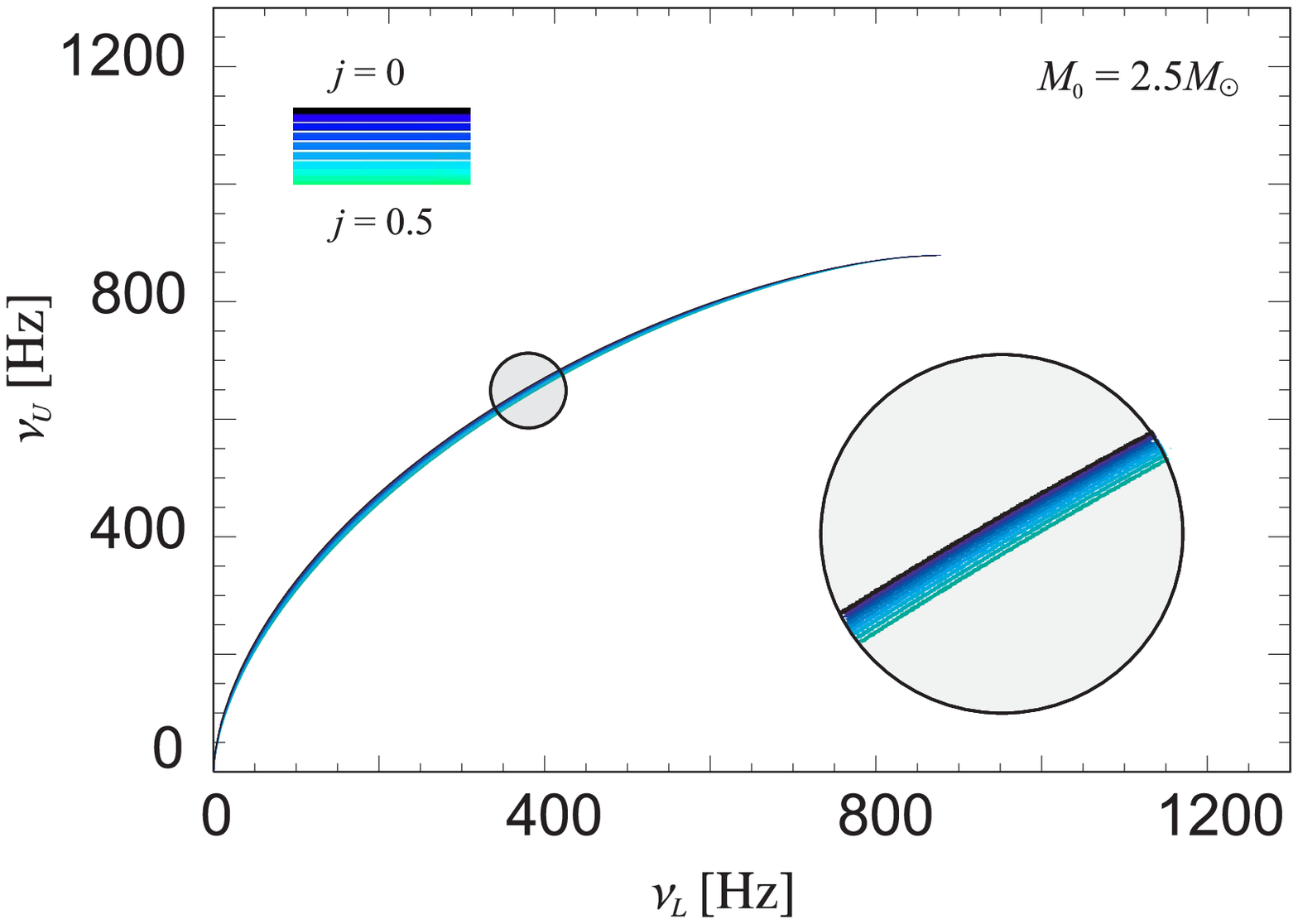}
\hfill
\includegraphics[width=.42\textwidth]{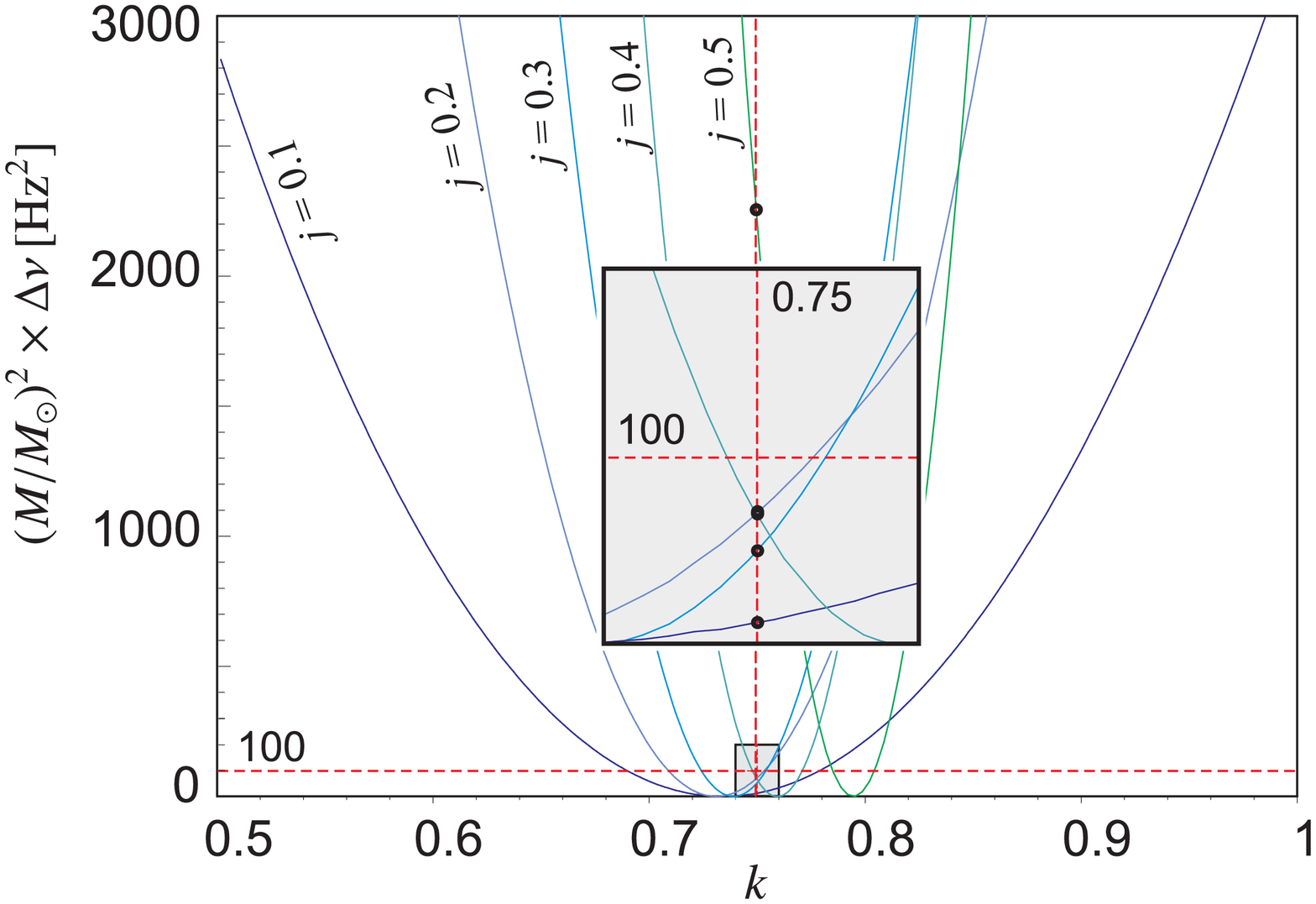}
\hfill~
\end{center}
\end{minipage}
\caption{Left: set of curves plotted for various combinations of $M$ and $j$ giving identical ISCO frequency. Right: the square of difference between the exact ISCO frequency and the frequency given by  Equation (\ref{equation:choice}).}
\label{figure:ISCO1}
\bigskip
\begin{minipage}{1\hsize}
\begin{center}
\hfill~
\includegraphics[width=.42\textwidth]{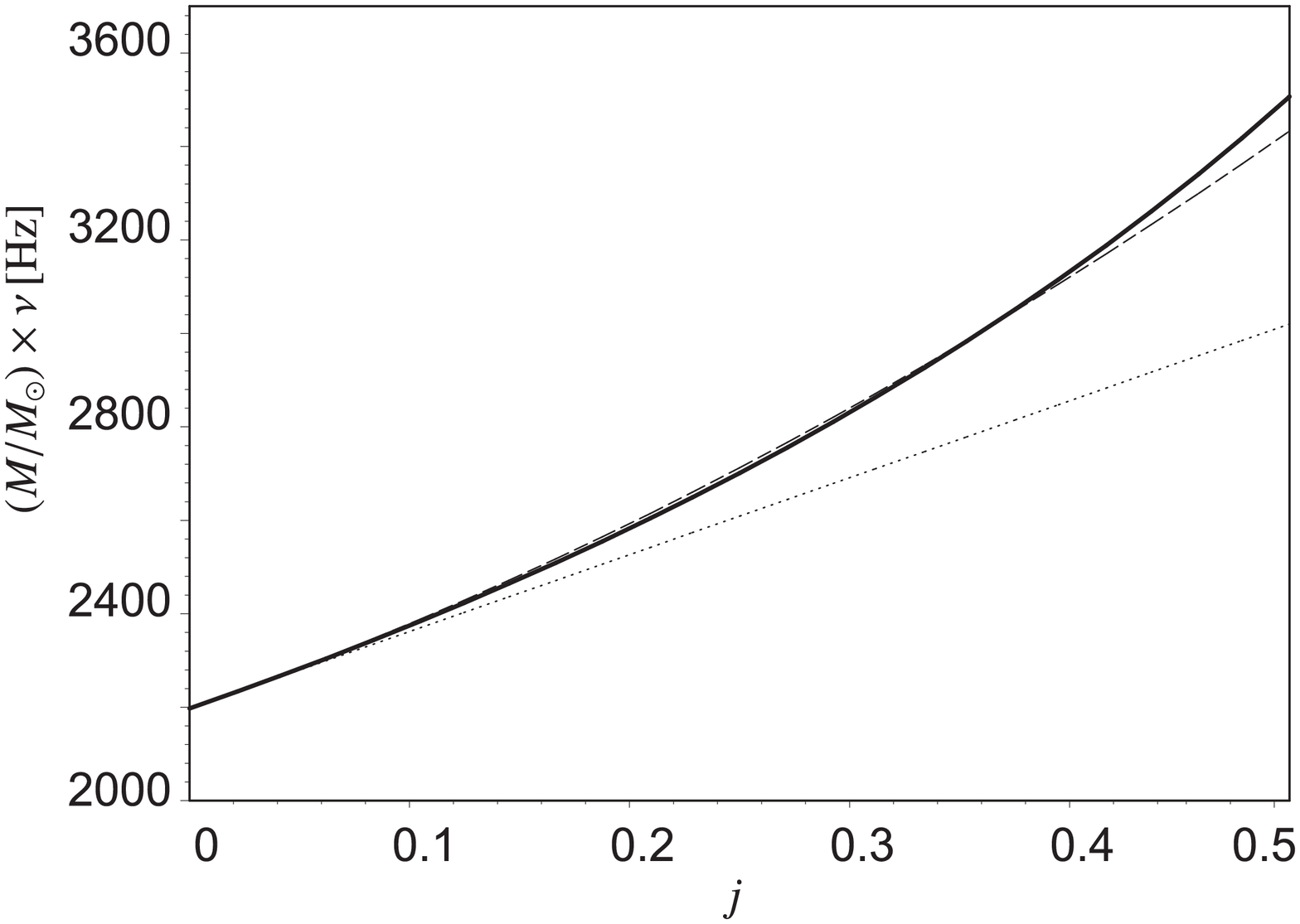}
\hfill
\includegraphics[width=.42\textwidth]{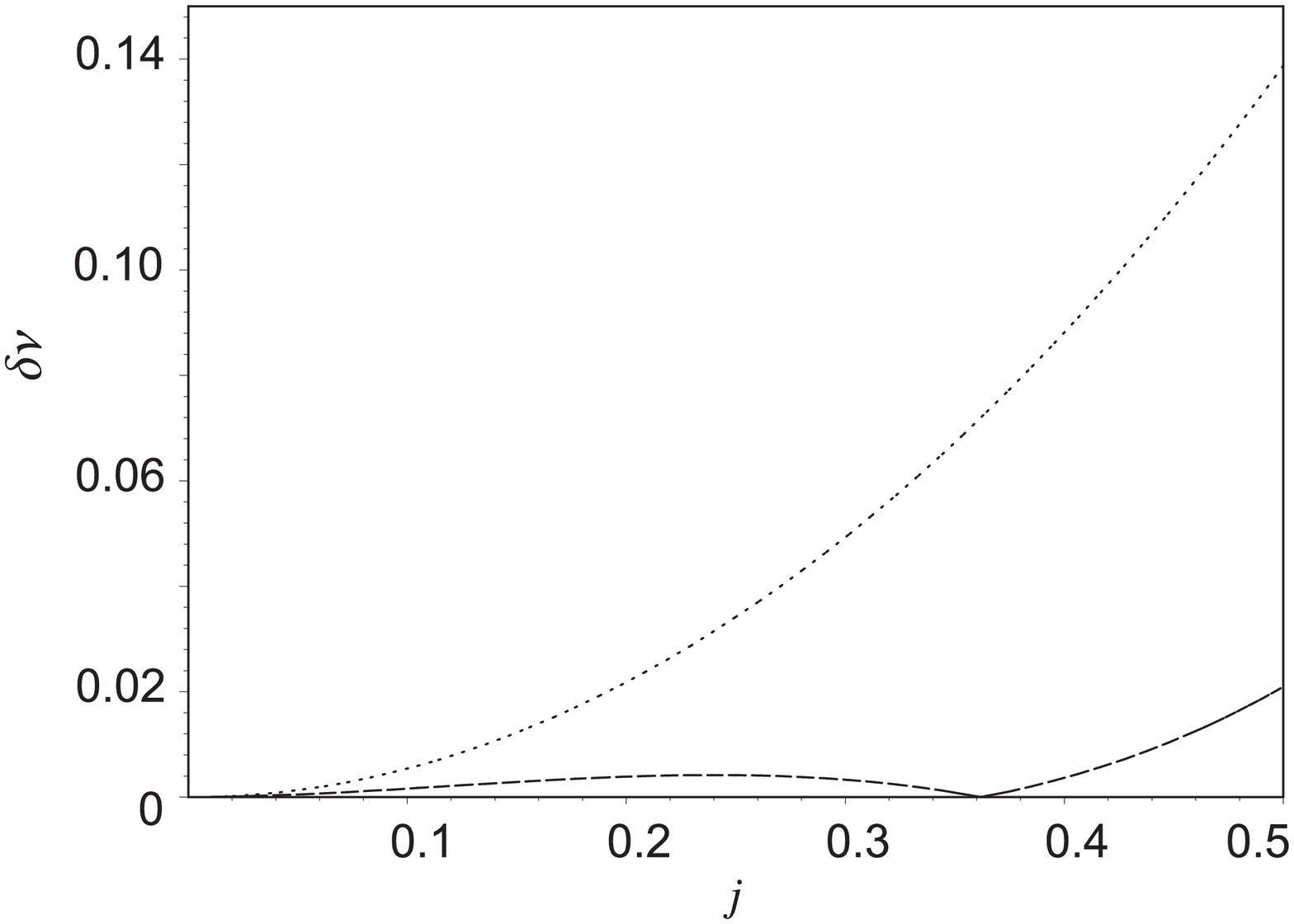}
\hfill~
\end{center}
\end{minipage}
\caption{\change{Left:  ISCO frequency calculated from Equation (\ref{equation:choice}) vs. exact relation implied by the Kerr solution (dashed vs. thick curve). The linear relation (\ref{equation:ISCO:kluzniak}) is shown as well for  comparison (dotted curve).  Right: the related relative difference from ISCO frequency in Kerr spacetime.}}
\label{figure:ISCO2}
\bigskip
\begin{minipage}{1\hsize}
\begin{center}
\hfill~
\includegraphics[width=.42\textwidth]{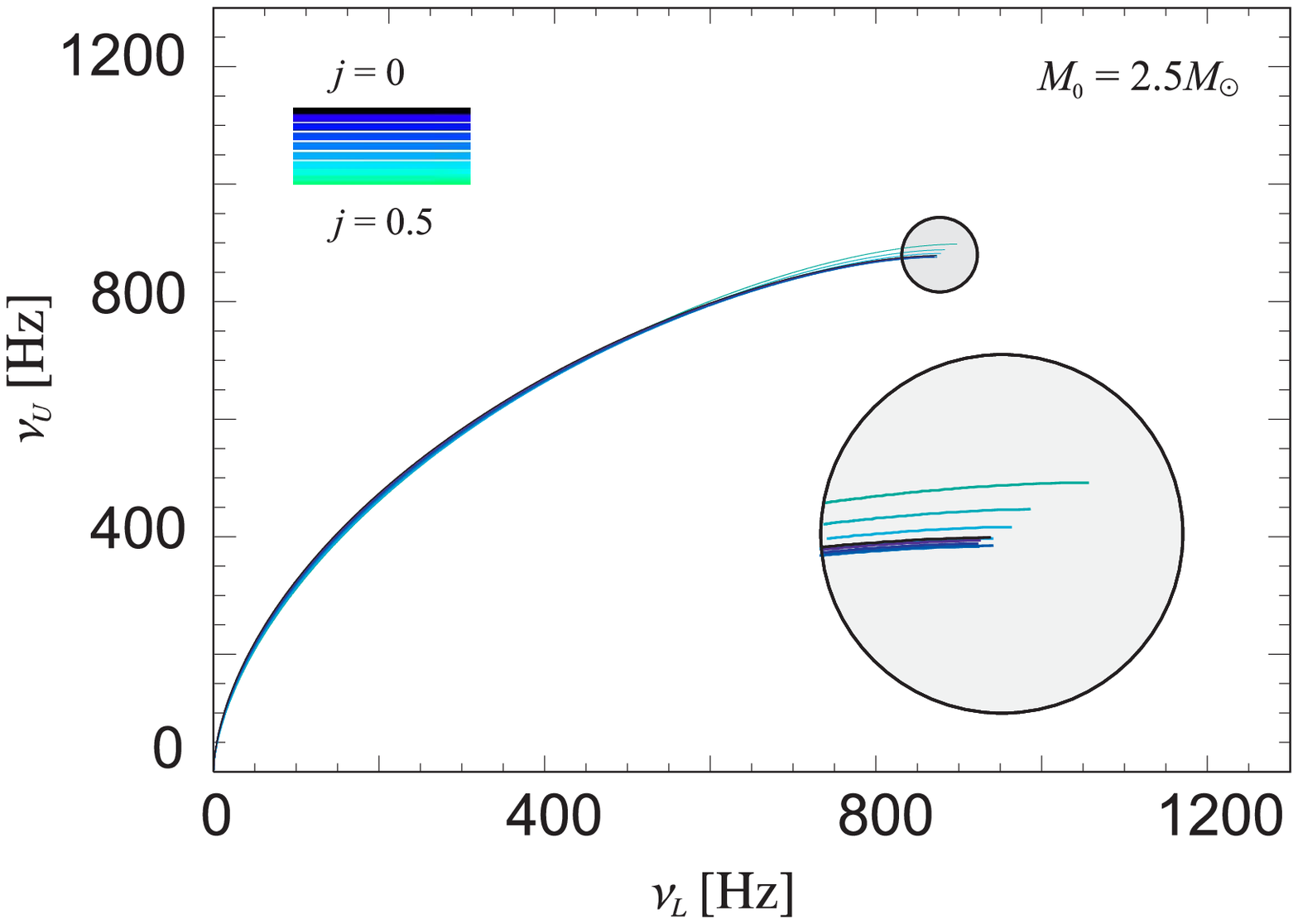}
\hfill
\includegraphics[width=.43\textwidth]{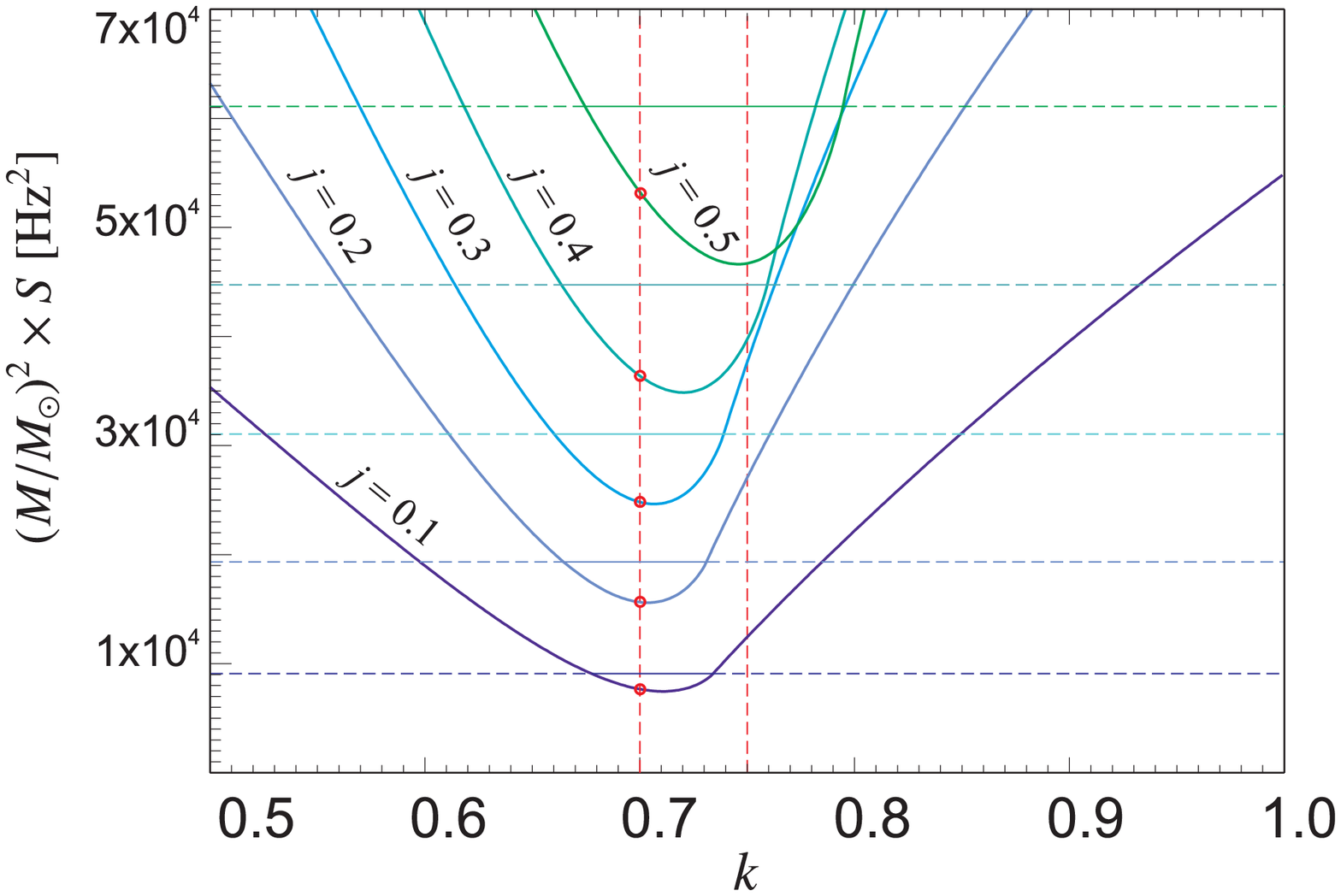}
\hfill~
\end{center}
\end{minipage}
\caption{Left: the set of curves plotted for combinations of $M$ and $j$ given by Equation (\ref{equation:ambiguity}) with $k=0.75$. Right: the integrated area $S$ related to Equation (\ref{equation:ambiguity}). Different values of $j$ are color-coded.  The same color-code is relevant for horizontal lines. These lines denote the values of $S$ arising for the set of curves numerically found from Equation (\ref{equation:rms}) and plotted in the left panel of Figure~\ref{figure:ISCO1}. The two red vertical lines denote the case of $k=0.75$ (curves $\nuU(\nuL)$ shown in the left panel of this figure) respectively $k=0.7$ (see the text for explanation).}
\label{figure:integral}
\end{figure*}

The ambiguity recognized in the previous section is implicitly given by dependence of the ISCO frequency on the NS angular momentum which for the Kerr metric follows from relations (\ref{equation:Keplerian:radial}) and (\ref{equation:rms}). In principle we can try to describe the ambiguity starting with these exact relations. The other option is to assume an approximative formula for the ISCO frequency. \change{One can expect that this formula should be at least of the second order in $j$ if consideration of spin up to $j=0.5$ is required. We check an arbitrarily simple form}
\begin{eqnarray}
\label{equation:choice}
\nonumber\\\nuK(x_\mathrm{ms})=(M_\sun/M)\times[1+k(j+j^2)]\times2197\,\mathrm{Hz}.
\end{eqnarray}
\change{The right panel of Figure \ref{figure:ISCO1} indicates the square of difference between the exact ISCO frequency in Kerr spacetimes following from Equation (\ref{equation:rms}) and the value following from Equation (\ref{equation:choice}). Inspecting the figure we can find that the particular choice of $k=0.75$ provides a very good approximation.}

\change{Figure~\ref{figure:ISCO2} then directly compares the exact relation and relation (\ref{equation:choice}) with $k=0.75$. For comparison, the first-order Taylor expansion formula (\ref{equation:ISCO:kluzniak}) is indicated.  Clearly, using Equation (\ref{equation:choice}) one may well approximate the Kerr-ISCO frequency up to $j\sim0.4$ and describe the discussed ambiguity in terms of {Schwarzschild} mass $M_0$ as}
\begin{eqnarray}
\label{equation:ambiguity}
M\sim[1+k(j+j^2)]\MS\,,
\end{eqnarray}
where $k=0.75$. In further discussion we therefore assume this formula.

\subsubsection{A.2.2. Comparison Between Curves}

The curves given by Equation (\ref{equation:ambiguity}) with $k=0.75$ are illustrated in the left panel of Figure~\ref{figure:integral}. Here we quantify their (apparent) conformity and investigate its dependence on $k$. It is natural to consider the integrated area $S$ between the curve for $M_0$, $j=0$ and the others as the relevant measure. The right panel of Figure~\ref{figure:integral} shows this area as the function of $k$ in Equation (\ref{equation:ambiguity}) for several values of $j$. The same panel also indicates the values related to the set of curves for mass found numerically from the exact Equations (\ref{equation:rms}), i.e., curves in the left panel of Figure~\ref{figure:ISCO1}. We can see that values of $S$ for $k=0.75$ are comparable to those related to Figure~\ref{figure:ISCO1}. Moreover, for a slightly different choice of $k=0.7$, all the values are smaller. {The ambiguity in mass with relation (\ref{equation:ambiguity}) is therefore best described for $k\sim0.7$ when the data uniformly cover the whole predicted curves.}

\begin{figure}[h!]
\begin{minipage}{1\hsize}
\begin{center}
\includegraphics[width=.85\textwidth]{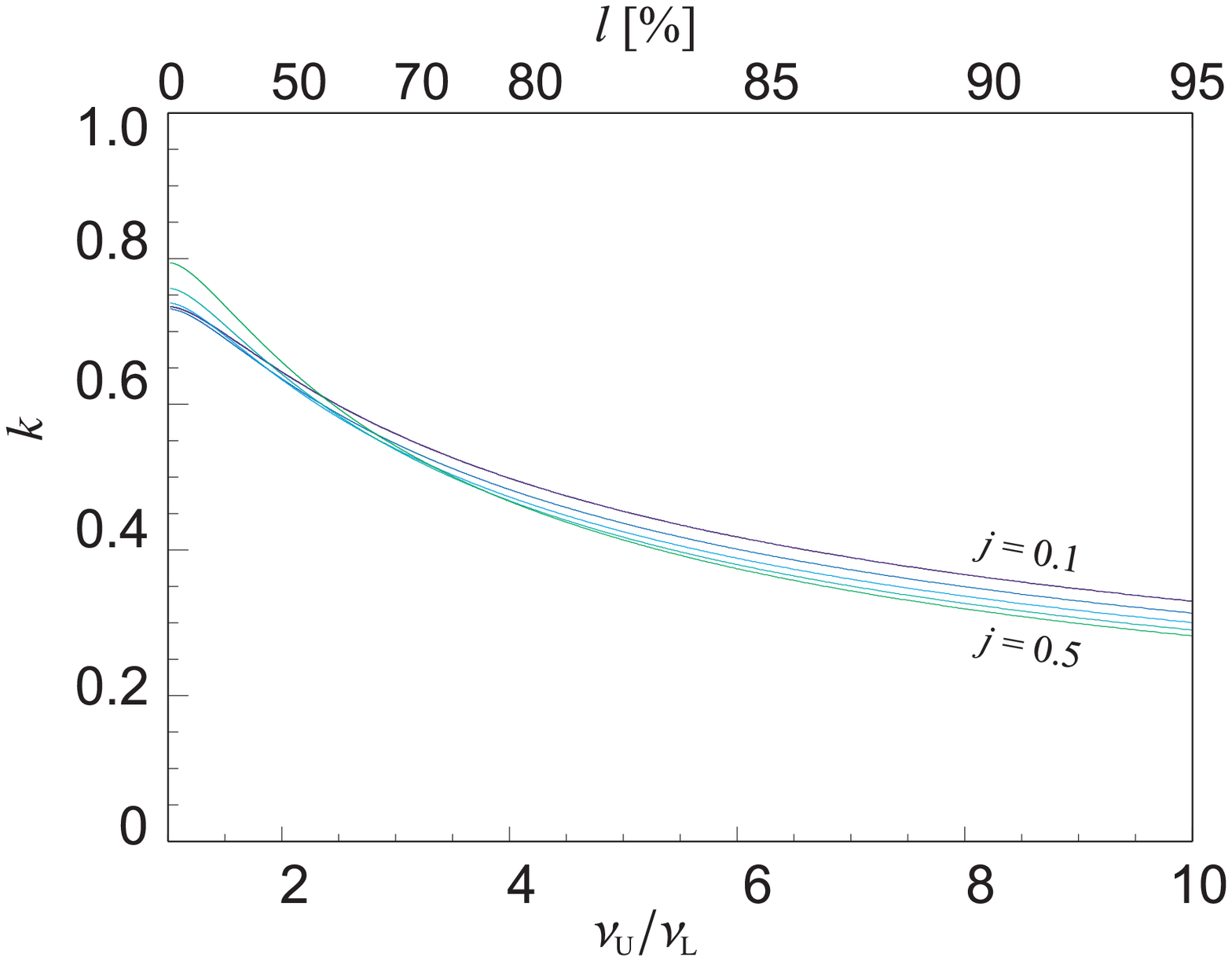}
\end{center}
\end{minipage}
\caption{Values of $k$ approximating the $M\fromto j$ ambiguity for the individual segments. The upper axes indicate the length of the curve $\nuU(\nuL)$ integrated from the ISCO point to the relevant frequency ratio.}
\label{figure:partial:points}
\end{figure}

The available data are restricted to certain frequency ranges and often exhibit clustering around some frequency ratios $\nuU/\nuL$ \citep[see][]{abr-etal:2003b, bel-etal:2007b, tor-etal:2008a, tor-etal:2008b, tor-etal:2008c,bar-bou:2008b,tor:2009,bou-etal:2009,bha:2009}. It is then useful to separately examine  the mass ambiguity for related segments of the curves. Such investigation is straightforward for small segments. Let us focus on a single point $[\nuL,~\nuU]$ representing a certain frequency ratio for a non-rotating star ($j=0$) of mass $M_0$. Assuming  relation (\ref{equation:ambiguity}) one may easily calculate the value of $k$ which rescales the mass to $M\neq M_0$ for a fixed non-zero $j$ in order to get exactly the same point $[\nuL,~\nuU]$. We applied this calculation for $\nuU/\nuL\in(1,~10)$ and $j\in(0,~0.5)$. The output is shown in Figure~\ref{figure:partial:points}. From the figure, it is possible to find $k$ that should best describe the ambiguity for a given frequency ratio (and thus for a small segment of data close to the ratio). It also indicates the length of the curve $\nuU(\nuL)$ integrated from the terminal (i.e., ISCO) point to the relevant frequency ratio (assuming $j=0$). This length is given in terms of the percentual share $l$ on the total length $\mathcal{L}$ of the curve $\nuU(\nuL)$, whereas the absolute numbers scale with $1/M$.

\begin{table*}[]
\begin{center}
\caption{The coefficient $k$ representing mass-angular momentum ambiguity (\ref{equation:ambiguity}).\label{table:k}}
\begin{tabular}{cccc}
\tableline\tableline
{segment}&$k$ in $M\sim[1+k(j+j^2)]\MS$&$l\,[\%]$&distance from ISCO $\times M_\sun/M$\,[km]\\
\tableline
$\nuL/\nuU\sim1.5$&0.75&25&\phantom{0}1\\
$\nuL/\nuU\sim2\phantom{.5}$&0.65&50&\phantom{0}3\\
$\nuL/\nuU\sim3\phantom{.5}$&0.55&70&\phantom{0}7\\
$\nuL/\nuU\sim4\phantom{.5}$&0.50&80&12\\
$\nuL/\nuU\sim5\phantom{.5}$&0.45&83&16\\
$\nuL/\nuU\sim6\phantom{.5}$&0.40&85&20\\
\tableline
{whole curve}&0.7$\phantom{0}$\\
\tableline
\end{tabular}
\end{center}
\end{table*}

Apparently, the segments with $\nuU/\nuL\in(1,~2)$ cover about 50$\%$ of the total length $\mathcal{L}$ while $k$ only slightly differs from the value of 0.7. We recall that this top part corresponds to the most of atoll and Z-sources data. For the segments of curves related to the sources exhibiting high frequency ratios such as Circinus X-1, there is an increasing deviation from the $0.7$ value and the coefficient reaches $k\sim0.6\fromto0.5$. A more detailed information is listed up to $\nuU/\nuL=5$ in Table \ref{table:k} providing the summary of this section.

\begin{thebibliography}{}


\bibitem[{Abramowicz \& Klu{\'z}niak}(2001)]{abr-klu:2001} Abramowicz, M. A., Klu{\'z}niak, W., 2001, A\&A, v.374, p.L19--L20
\bibitem[{Abramowicz et al.}(2003a)]{abr-etal:2003a} Abramowicz, M. A., Almergren, G. J. E., Klu{\'{z}}niak, W., \& Thampan, A. V., 2003a, arXiv: gr-qc/0312070
\bibitem [{Abramowicz et al.}(2003b)]{abr-etal:2003b} Abramowicz, M. A., Bulik, T., Bursa, M., Klu{\'z}niak, W., 2003a, A\&A, 404, L21-L24
\bibitem [{Abramowicz et al.}(2003c)]{abr-etal:2003c} Abramowicz, M. A., Karas, V., Klu{\'z}niak, W, Lee, W. H., Rebusco, P., 2003b, PASJ, 55, 466-467
\bibitem[{Akmal et al.}(1998)]{akm-etal:1998} Akmal, A., {Pandharipande}, V. R., {Ravenhall}, D. G., 1998, {\prc}, 58, 1804-1828, arXiv:hep-ph/9804388
\bibitem[{Aliev \& Galtsov}(1981)]{ali-gal:1981} Aliev, A. N., Galtsov, D. V., 1981, GRG, 13, 899
\bibitem[{Alpar \& Shaham}(1985)]{alp-sha:1985} Alpar, M. A., Shaham, J., 1985, Nature, 316, 239 
\bibitem [{Bardeen et al.}(1972)] {bar-etal:1972} Bardeen, J. M., Press, W. H., \& Teukolsky, S. A., 1972, ApJ, 178, 347-370
\bibitem[{Barret {et~al.}}(2006)]{bar-etal:2006} Barret, D., Olive, J. F., Miller, M. C., 2006, Monthly Notices of the Royal Astronomical Society, 370, 3, 1140-1146
\bibitem[{Barret et~al.}(2005a)]{bar-etal:2005a}
Barret, D., Olive, J. F., Miller, M. C., 2005a, MNRAS, 361, 855
\bibitem[{Barret et~al.}(2005b)]{bar-etal:2005b}
Barret, D., Olive, J. F., Miller, M. C., 2005b, Astronomical Notes, 326, 9, 808
\bibitem[{Barret \& Boutelier}(2008a)]{bar-bou:2008a} Barret, D., Boutelier, M., 2008a, MNRAS, 384, 4, 1519-1524
\bibitem[{Barret \& Boutelier}(2008b)]{bar-bou:2008b} Barret, D., Boutelier, M., 2008b, New Astronomy Reviews, 51, 10-12, 835-840
\bibitem[{Belloni et~al.}(2005)]{bel-etal:2005} Belloni, T., M{\'e}ndez, M., Homan, J., 2005, Astronomy and Astrophysics,
  437, 209
\bibitem[{Belloni et al.}(2007a)]{bel-etal:2007a} Belloni, T., M\'endez, M., Homan, J., 2007a, MNRAS, 376, 3, 1133-1138
\bibitem[{Belloni et al.}(2007b)]{bel-etal:2007b} Belloni, T., M\'endez, M., Homan, J., 2007b, MNRAS, 379, 1, 247, e-print arXiv:0705.0793
\bibitem[{Berti et al.}(2005)]{ber-etal:2005} Berti, E., White, F., Maniopoulou, A., Bruni, M., 2005, Monthly Notices of the Royal Astronomical Society, Volume 358, Issue 3, pp. 923-938
\bibitem[{Bhattacharyya}(2009)]{bha:2009} Bhattacharyya, S., 2009, 2009, Res. Astron. Astrophys., 10, 227, e-print arXiv:0911.5574
\bibitem[{Blaes et al.}(2007)]{bla-etal:2007} Blaes, O. M., \v{S}r\'{a}mkov\'{a}, E., Abramowicz, M. A., Klu\'zniak, W., Torkelsson, U., 2007,	The Astrophysical Journal, Volume 665, Issue 1, pp. 642-653
\bibitem[{Boirin et al.}(2000)]{Boirin-00}
Boirin, L., Barret, D., Olive, J.F., Bloser, P.F., Grindlay, J.E.,
2000, A\&A, 361, 121
\bibitem[{Boutelier et al.}(2009)]{bou-etal:2009} Boutelier, M., Barret, D., Lin, Y., T\"{o}r\"{o}k, G., 2009, MNRAS, volume 401, Issue 2, pp. 1290-1298e-print arXiv:0909.2990
\bibitem[{Boutloukos et al.}(2006)]{bou-etal} Boutloukos, S., van der Klis, M., Altamirano, D., Klein-Wolt, M., Wijnands, R., Jonker, P. G., Fender, R. P., 2006, ApJ, Volume 653, Issue 2, 1435-1444; also Volume 664, Issue 1, pp. 596-596
\bibitem[{Boyer \& Lindquist}(1967)]{boy-lin:1967} Boyer, R. H., Lindquist, R. W., 1967, J. Math. Phys. 8: 265–281
\bibitem[{Chandrasekhar \& Miller}(1974)] {cha-mil:1974} Chandrasekhar, S. , Miller, J.~C., MNRAS, 1974, 167, p. 63-80
\bibitem[{Carter}(1971)]{car:1971} Carter, B., 1971, Physical Review Letters 26, 331–333
\bibitem[{\v{Ca}de\v{z} et al.}(2008)]{cad-etal:2008} \v{Ca}de\v{z}, A., Calvani, M., Kosti\'{c}, U., 2008, Astronomy and Astrophysics, Volume 487, Issue 2,  pp.527-532
\bibitem[{Cook et al.}(1994)]{coo-etal:1994} Cook, G. B., Shapiro, S. L., Teukolsky, S. A., 1994, Astrophysical Journal, Part 1 (ISSN 0004-637X), vol. 422, no. 1, p. 227-242
\bibitem[{di Salvo et al.}(2003)]{diSalvo-03}
Di Salvo, T., M\'endez, M., van der Klis, M., 2003, A\&A, 406, 177
\bibitem [{Gourgoulhon et al.}(2000)]{LORENE} Gourgoulhon, E., Grandcl\'ement, P., Marck, J.A., Novak, J., Taniguchi, K., 2000, http://www.lorene.obspm.fr/
\bibitem[{Hartle}(1967)]{har:1967} Hartle, J. B., 1967, Astrophysical Journal, vol. 150, p.1005
\bibitem[{Hartle \& Sharp}(1967)]{har-sha:1967} Hartle, J. B., Sharp, D. V., 1967, Astrophysical Journal, vol. 147, p.317
\bibitem[{Hartle \& Thorne}(1968)]{har-tho:1968}	Hartle, J. B., Thorne, K. S., Astrophysical Journal, vol. 153, p.807
\bibitem[{Hor\'ak}(2008)]{hor:2008} Hor\'ak, J., Astronomy and Astrophysics, Volume 486, Issue 1, 2008, pp.1-8
\bibitem[{Hor\'ak et al.}(2009)]{hor-etal:2009} Hor\'ak, J., Abramowicz, M. A., Klu{\'z}niak, W., Rebusco, P., Török, G., Astronomy and Astrophysics, Volume 499, Issue 2, 2009, pp.535-540
\bibitem[{Homan et al.}(2002)]{Homan-02} 
Homan, J., van der Klis, M., Jonker, P.G., Wijnands, R., Kuulkers, E.,
M\'endez, M., Lewin, W.H.G., 2002, ApJ, 568, 878
\bibitem[{Jonker et al.}(2002a)]{Jonker2002a} Jonker, P. G., van der Klis, M., Homan, J., M\'endez, M., Lewin, W. H. G., Wijnands, R., Zhang, W., 2002a, MNRAS, 333, 665
\bibitem[{Jonker et al.}(2002b)]{Jonker2002b} Jonker, P. G., M\'endez, M., van der Klis, M., 2002b, MNRAS, 336, L1
\bibitem[{Karas}(1999)]{kar:1999} Karas, V., ApJ, Volume 526, Issue 2, 1999, pp. 953-956
\bibitem[{Kato et al.}(1998)]{kat-etal:1998} Kato,S., Fukue, J., Mineshige, S., Black-hole accretion disks. Edited by Shoji Kato, Jun Fukue, and Sin Mineshige. Publisher: Kyoto, Japan: Kyoto University Press, 1998. ISBN: 4876980535
\bibitem[{Kato}(2009a)]{kat:2009a} Kato, S., 2009a, Publications of the Astronomical Society of Japan, Vol.61, No.6, pp.1237--1245
\bibitem[{Kato}(2009b)]{kat:2009b} Kato, S., 2009b, Publications of the Astronomical Society of Japan, Vol.60, No.4, pp.889--897
\bibitem[{Kato}(2008)]{kat:2008} Kato, S., 2008, Publications of the Astronomical Society of Japan, Vol.60, No.1, pp.111--123
\bibitem[{Kato}(2007)]{kat:2007} Kato, S., 2007, Publications of the Astronomical Society of Japan, Vol.59, No.2, pp.451-455
\bibitem[{Kerr}(1963)]{ker:1963} Kerr, R. P., 1963, Physical Review Letters 11: 237–238
\bibitem[{van~der Klis}(2006)]{Kli:2006:CompStelX-Ray:}
van~der Klis, M., in {Compact Stellar X-Ray Sources}, ed. W.~H.~G. Lewin
  \& M.~van~der Klis (Cambridge: Cambridge University Press), 2006, 39--112; see also astro-ph/0410551
\bibitem[{Klu\'zniak et al.}(1990)]{klu-etal:1990} Klu\'zniak, W., Michelson, P., Wagoner, R. V., 1990, ApJ, Part 1 (ISSN 0004-637X), 358, Aug. 1, 538-544
\bibitem[{Klu\'zniak \& Wagoner}(1985)]{klu-wag:1985} Klu\'zniak, W., Wagoner, R. V., 1985, ApJ, Part 1 (ISSN 0004-637X), vol. 297, Oct. 15, 548-554
\bibitem[{Klu{\'z}niak \& Abramowicz}(2001)]{klu-abr:2001} Abramowicz, M. A., Klu{\'z}niak, W., eprint arXiv:astro-ph/0105057
\bibitem[{Klu\'zniak et al.}(2004)]{klu-etal:2004} Klu\'zniak, W., Abramowicz, M. A., Kato, S., Lee, W. H., Stergioulas, N., 2004, The Astrophysical Journal, Volume 603, Issue 2, pp. L89-L92
\bibitem[{Klu\'zniak}(2008)]{klu:2008} Klu\'zniak, W., 2008, New Astronomy Reviews, Volume 51, Issue 10-12, p. 841-845
\bibitem[{Komatsu et al.}(1989)] {kom-etal:1989} Komatsu, H., Eriguchi, Y., Hachisu, I., 1989,	Royal Astronomical Society, Monthly Notices (ISSN 0035-8711), vol. 237, March 15, 355-379
\bibitem[{Kosti\'{c} et al.}(2009)]{kos-etal:2009} Kosti\'{c}, U., \v{C}ade\v{z}, A., Calvani, M., Gomboc, A., 2009, Astronomy and Astrophysics, Volume 496, Issue 2, pp.307-315
\bibitem[{Lamb et al.}(1985)]{lam-etal:1985} Lamb, F. K., Shibazaki, N., Alpar, M. A., Shaham, J., 1985, Nature, 317, 681
\bibitem[{Lamb \& Coleman}(2001)]{lam-col:2001} Lamb, F. K., Miller, M. C., 2001, The Astrophysical Journal, Volume 554, Issue 2, pp. 1210-1215
\bibitem[{Lamb \& Coleman}(2003)]{lam-col:2003} Lamb, F. K., Miller, M. C., eprint arXiv:astro-ph/0308179
\bibitem[{Lamb \& Boutloukos}(2007)]{Lam-Bou:2007:ASSL:ShrtPerBS}
 Lamb, F. K., \& Boutloukos, S., in {Astrophysics and Space Science
  Library}, Vol. 352, {Short-Period Binary Stars: Observations, Analyses, and
  Results}, ed. E.~F. Milone, D.~A. Leahy, \& D.~Hobill (Dordrecht: Springer), 2007
\bibitem[{Lattimer \& Prakash}(2001)]{lat-pra:2001} Lattimer, J. M., Prakash, M., 2001, \apj, vol. 550, p. {426-442}, {arXiv:astro-ph/0002232},
\bibitem[{Lattimer \& Prakash}(2007)]{lat-pra:2007} Lattimer, J. M., Prakash, M., 2007, \physrep, vol. 442, p. {109-165}, {arXiv:astro-ph/0612440}
\bibitem[{Lee \& Miller}(1998)]{lee-mil:1998}	Lee, H. C., Miller, G. S., 1989, Monthly Notices of the Royal Astronomical Society, Volume 299, Issue 2, pp. 479-487
\bibitem[{Meheut \& Tagger}(2009)]{meh-tag:2009} Meheut, H., Tagger, M., 2009, Monthly Notices of the Royal Astronomical Society, Volume 399, Issue 2, pp. 794-800
\bibitem[{M\'endez \& van der Klis}(2000)]{MvdK2000} M\'endez, M., van der Klis, M., 2000, MNRAS, 318, 938
\bibitem[{M\'endez et al.}(2001)]{M2001} M\'endez, M., van der Klis, M., Ford, E.C., 2001, ApJ, 561, 1016
\bibitem[{Miller}(1977)]{mil:1977} Miller, J.~C., MNRAS, 1977, 179, p. 483-498
\bibitem[{Miller at al.}(1998a)]{mil-etal:1998a} Miller, M. C., Lamb, F. K., Psaltis, D., 1998b, The Astrophysical Journal, 508, 791
\bibitem[{Miller et al.}(1998b)]{mil-etal:1998b} Miller, M. C, Lamb, F. K., \& Cook, G. B., 1998a, ApJ, 509, 793 
\bibitem[{Miller}(2006)]{mil:2006}	Miller, M. C., 2006, Advances in Space Research, Volume 38, Issue 12, p. 2680-2683.
\bibitem[{Morsink \& Stella}(1999)]{mor-ste:1999} Morsink, S. M., Stella, L., 1999, The Astrophysical Journal, Volume 513, Issue 2, pp. 827-844
\bibitem[{Mukhopadhyay}(2009)]{muk:2009} Mukhopadhyay, B., 2009, ApJ, 694, 387
\bibitem[{Nozawa et al.}(1998)]{noz-etal:1998} Nozawa, T., Stergioulas, N., Gourgoulhon, E., Eriguchi, Y., 1998, Astronomy and Astrophysics Supplement, v.132, p.431-454 
\bibitem [{Okazaki et al.}(1987)]{oka-etal:1987} Okazaki, A. T., Kato, S., \& Fukue, J., 1987, Publ. Astron. Soc. Japan, 39, 457
\bibitem[{Pach\'on et al.}(2006)]{pac-etal:2006} Pach\'on, L. A., Rueda, J. A., Sanabria-G\'omez, J. D., 2006, Physical Review D, vol. 73, Issue 10, id. 104038 
\bibitem[{P\'etri}(2005a)]{pet:2005a} P\'etri, J., 2005a, Astronomy and Astrophysics, Volume 439, Issue 2, pp.L27-L30
\bibitem[{P\'etri}(2005b)]{pet:2005b} P\'etri, J., 2005b, Astronomy and Astrophysics, Volume 439, Issue 2, pp.443-459
\bibitem[{P\'etri}(2005c)]{pet:2005c} P\'etri, J., 2005c, Astronomy and Astrophysics, Volume 443, Issue 3, pp.777-780
\bibitem[{Psaltis et al.}(1999)]{psa-etal:1999} Psaltis, D., Wijnands, R., Homan, J., Jonker, P. G., van der Klis, M., Miller, M. C., Lamb, F. K., Kuulkers, E., van Paradijs, J., Lewin, W. H. G., 1999,	The Astrophysical Journal, Volume 520, Issue 2, pp. 763-775
\bibitem[{Press et al.} (2007)]{pre-etal:2007} Press, W. H., Teukolsky, S. A., Vetterling, W. T., Flannery, B. P., 2007, Numerical Recipes: The Art of Scientific Computing, Third Edition, Cambridge University Press
\bibitem[{Rezzolla et al.}(2003)]{rez-etal:2003} Rezzolla, L., Yoshida, S., Zanotti, O., 2003,	Monthly Notices of the Royal Astronomical Society, Volume 344, Issue 3, pp. 978-992
\bibitem[{Rezzolla}(2004)]{rez:2004} Rezzolla, L., 2004, X-RAY TIMING 2003: Rossie and Beyond. AIP Conference Proceedings, Volume 714, pp. 36-39
\bibitem[{Rikovska Stone et al.}(2003)]{rik-etal:2003} Rikovska Stone, J., Miller, J. C., Koncewicz, R., Stevenson, P. D., Strayer, M. R., 2003, {\prc}, vol. 68, n. 3
\bibitem[{Schnittman \& Rezzolla}(2006)]{sch-rez:2006} Schnittman, J. D., Rezzolla, L., 2006, The Astrophysical Journal, Volume 637, Issue 2, pp. L113-L116
\bibitem[{Shibata \& Sasaki}(1998)]{shi-sas:1998} Shibata, M., and Sasaki, M., 1998, Phys. Rev. D 58, 104011
\bibitem[{Straub \& \v{S}r\'{a}mkov\'{a}}(2009)]{str-sra:2009}	Straub, O., \v{S}r\'{a}mkov\'{a}, E., 2009, Classical and Quantum Gravity, Volume 26, Issue 5, pp. 055011
\bibitem[{\v{S}r\'{a}mkov\'{a}}(2005)]{sra:2005}	\v{S}r\'{a}mkov\'{a}, E., 2005, Astronomische Nachrichten, Vol.326, Issue 9, p.835-837 
\bibitem[{\v{S}r\'{a}mkov\'{a} et al.}(2007)]{sra-etal:2007}	\v{S}r\'{a}mkov\'{a}, E., Torkelsson, U., Abramowicz, M. A., 2007, Astronomy and Astrophysics, Volume 467, Issue 2, pp.641-646
\bibitem[{Stella \& Vietri}(1998)]{ste-vie:1998}
Stella, L., Vietri, M., 1998, in Abstracts of the 19th Texas Symposium on Relativistic Astrophysics and Cosmology, ed. J. Paul, T.
Montmerle \& E. Aubourg (CEA Saclay) 
\bibitem[{Stella \& Vietri}(1999)]{ste-vie:1999}
Stella, L., Vietri, M., 1999, \prl, 82, 17.
\bibitem[{Stella \& Vietri}(2002)]{ste-vie:2002} Stella, L., Vietri, M., 2002, in The Ninth Marcel Grossmann Meeting. Proceedings of the MGIXMM Meeting held at The University of Rome "La Sapienza", 2-8 July 2000, ed. V. G. Gurzadyan, R. T. Jantzen \& R. Ruffini (World Scientific Publishing), Part A, 426
\bibitem[{Stella et al.}(1999)]{ste-etal:1999} Stella, L., Vietri, M., Morsink, S. M., 1999, The Astrophysical Journal, Volume 524, Issue 1, pp. L63-L66
\bibitem[{Stergioulas \& Friedman}(1995)]{ste-fri:1995} Stergioulas, N., Friedman, J. L., 1995,	Astrophysical Journal, Part 1 (ISSN 0004-637X), vol. 444, no. 1, p. 306-311
\bibitem[{Stergioulas \& Morsink}(1997)]{RNS} Stergioulas, N., Morsink, S. M., 1997, http://www.gravity.phys.uwm.edu/rns/
\bibitem[{van Straaten et al.}(2000)]
{van-Straaten-00}
van Straaten, S., Ford, E.C., van der Klis, M., M\'endez, M., Kaaret, P., 2000,
ApJ, 540, 1049
\bibitem[{van Straaten et al.}(2002)] {van-Straaten-02} van Straaten, S., van der Klis, di Salvo, T., Belloni, T., 2002, ApJ, 568, 912
\bibitem[{Stuchl\'{\i}k et al.}(2008)]{stu-etal:2008} Stuchl\'{\i}k, Z., Konar, S., Miller, J. C., S. Hled\'{\i}k, 2008, Astronomy and Astrophysics, Volume 489, Issue 3, 2008, pp.963-966
\bibitem[{Thirring \& Lense}(1918)]{thi-len:1918} Thirring, H., Lense, J., 1918, Phys. Z. 19, 156
\bibitem[{Titarchuk \& Kent}(2002)] {tit-ken:2002}	Titarchuk, L., Kent, W.,	2002, The Astrophysical Journal, Volume 577, Issue 1, pp. L23-L26
\bibitem[{Titarchuk}(2002)] {tit:2002}	Titarchuk, L., 2002, The Astrophysical Journal, Volume 578, Issue 1, pp. L71-L74, e-print	arXiv:astro-ph/0208423v1
\bibitem[{T\"or\"ok \& Stuchl{\'{\i}}k}(2005)]{tor-stu:2005} T\"or\"ok, G. , Stuchl{\'{\i}}k, Z., 2005, Astronomy and Astrophysics, Volume 437, Issue 3, pp.775-788
\bibitem[{T\"or\"ok et al.}(2008a)]{tor-etal:2008a} T\"or\"ok, G., Bakala, P., Stuchl{\'{\i}}k, Z., \v{C}ech, P., 2008a, Acta Astronomica, 58/1, pp. 1--14
\bibitem[{T\"{o}r\"{o}k et~al.}(2008b)]{tor-etal:2008b} T{\"{o}}r{\"{o}}k, G., Abramowicz, M. A., Bakala, P., Bursa, M., Hor\'{a}k, J., Klu{\'z}niak, W., Rebusco, P., Stuchl\'{\i}k, Z., 2008b, Acta Astronomica, 58/1, pp. 15--21, arXiv:0802.4070
\bibitem[{T\"{o}r\"{o}k et~al.}(2008c)]{tor-etal:2008c} T{\"{o}}r{\"{o}}k, G., Abramowicz, M. A., Bakala, P., Bursa, M., Hor\'{a}k, J., Rebusco, P., Stuchl\'{\i}k, Z., 2008c, Acta Astronomica, 58/2, arXiv:0802.4026 
\bibitem[{T\"{o}r\"{o}k}(2009)]{tor:2009} T{\"{o}}r{\"{o}}k, G., 2009, A\&A, vol. 497, 3, p. 661-665, arXiv:0812.4751
\bibitem[{Weber}(1999)]{web:1999} Weber, F., 1999, Pulsars as astrophysical laboratories for nuclear and particle physics, Bristol, U.K., Institute of Physics, QB 464 W42 1999. DA
\bibitem[{Wiringa et al.}(1988)]{wir-etal:1988} Wiringa, R. B., {Fiks}, V., {Fabrocini}, A., 1988, {\prc}, vol. 38, p. {1010-1037}
\bibitem[{Yan et al.}(2009)]{yan-etal:2009} Yan, C. M., Zhang, Y., Yin, H. X., Yan, Zhao, Y. H., 2009,  Astronomische Nachrichten, Vol.330, Issue 4, p.398    
\bibitem[{Zhang et al.}(2010)]{zha-etal:2010} Zhang, C.M., Wei, Y.C., Yin, H.X., Zhao, Y.H., Lei, Y.J., Song, L.M., Zhang, F., Yan, Y., 2010, Sci. China - Phys. Mech. \& Astron, 53 (Suppl.1), 114 (arXiv:0912.0768)
\bibitem[{Zhang et al.}(1998)]{Zhang-98}
Zhang, W., Smale, A.P., Strohmayer, T.E., Swank, J.H., 1998, ApJ, 500, L171
\bibitem[{Zhang et~al.}(2006)]{Zha-etal:2006:MONNR:kHzQPOFrCorr} Zhang, C. M., Yin, H. X., Zhao, Y. H., Zhang, F., \& Song, L. M., 2006, Monthly  Notices Roy. Astronom. Soc., 366, 1373
\bibitem[{Zhang et al.}(2007a)]{zha-etal:2007a} Zhang, C. M., Yin, H. X., Zhao, Y. H., 2007a, The Publications of the Astronomical Society of the Pacific, Volume 119, Issue 854, pp. 393-397
\bibitem[{Zhang et al.}(2007b)]{zha-etal:2007b} Zhang, C. M., Yin, H. X., Zhao, Y. H., Song, L. M., 2007b, Astronomische Nachrichten, Vol.328, Issue 6, p.491
\bibitem[{Zhang}(2005)]{zha:2005} Zhang, C. M., 2005, Chinese Journal of Astronomy \& Astrophysics, Vol. 5 Supplement, p. 21-26

  
\end{thebibliography}
\end{document}